\PassOptionsToPackage{unicode}{hyperref}
\PassOptionsToPackage{hyphens}{url}
\PassOptionsToPackage{dvipsnames,svgnames,x11names}{xcolor}
\documentclass[
  letterpaper,
  DIV=11,
  numbers=noendperiod]{scrartcl}
\usepackage{xcolor}
\usepackage[margin=1in]{geometry}
\usepackage{amsmath,amssymb}
\usepackage{iftex}
\ifPDFTeX
  \usepackage[T1]{fontenc}
  \usepackage[utf8]{inputenc}
  \usepackage{textcomp} 
\else 
  \usepackage{unicode-math} 
  \defaultfontfeatures{Scale=MatchLowercase}
  \defaultfontfeatures[\rmfamily]{Ligatures=TeX,Scale=1}
\fi
\usepackage{lmodern}
\ifPDFTeX\else
\fi
\IfFileExists{upquote.sty}{\usepackage{upquote}}{}
\IfFileExists{microtype.sty}{
  \usepackage[]{microtype}
  \UseMicrotypeSet[protrusion]{basicmath} 
}{}
\usepackage{setspace}
\makeatletter
\@ifundefined{KOMAClassName}{
  \IfFileExists{parskip.sty}{%
    \usepackage{parskip}
  }{
    \setlength{\parindent}{0pt}
    \setlength{\parskip}{6pt plus 2pt minus 1pt}}
}{
  \KOMAoptions{parskip=half}}
\makeatother
\makeatletter
\ifx\paragraph\undefined\else
  \let\oldparagraph\paragraph
  \renewcommand{\paragraph}{
    \@ifstar
      \xxxParagraphStar
      \xxxParagraphNoStar
  }
  \newcommand{\xxxParagraphStar}[1]{\oldparagraph*{#1}\mbox{}}
  \newcommand{\xxxParagraphNoStar}[1]{\oldparagraph{#1}\mbox{}}
\fi
\ifx\subparagraph\undefined\else
  \let\oldsubparagraph\subparagraph
  \renewcommand{\subparagraph}{
    \@ifstar
      \xxxSubParagraphStar
      \xxxSubParagraphNoStar
  }
  \newcommand{\xxxSubParagraphStar}[1]{\oldsubparagraph*{#1}\mbox{}}
  \newcommand{\xxxSubParagraphNoStar}[1]{\oldsubparagraph{#1}\mbox{}}
\fi
\makeatother

\usepackage{longtable,booktabs,array}
\usepackage{calc} 
\usepackage{etoolbox}
\makeatletter
\patchcmd\longtable{\par}{\if@noskipsec\mbox{}\fi\par}{}{}
\makeatother
\IfFileExists{footnotehyper.sty}{\usepackage{footnotehyper}}{\usepackage{footnote}}
\makesavenoteenv{longtable}
\usepackage{graphicx}
\makeatletter
\newsavebox\pandoc@box
\newcommand*\pandocbounded[1]{
  \sbox\pandoc@box{#1}%
  \Gscale@div\@tempa{\textheight}{\dimexpr\ht\pandoc@box+\dp\pandoc@box\relax}%
  \Gscale@div\@tempb{\linewidth}{\wd\pandoc@box}%
  \ifdim\@tempb\p@<\@tempa\p@\let\@tempa\@tempb\fi
  \ifdim\@tempa\p@<\p@\scalebox{\@tempa}{\usebox\pandoc@box}%
  \else\usebox{\pandoc@box}%
  \fi%
}
\def\fps@figure{htbp}
\makeatother

\NewDocumentCommand\citeproctext{}{}

\makeatletter
 \let\@cite@ofmt\@firstofone
 \def\@biblabel#1{}
 \def\@cite#1#2{{#1\if@tempswa , #2\fi}}
\makeatother
\newlength{\cslhangindent}
\newlength{\csllabelwidth}
\newenvironment{CSLReferences}[2] 
 {\begin{list}{}{%
  \setlength{\itemindent}{0pt}
  \setlength{\leftmargin}{0pt}
  \setlength{\parsep}{0pt}
  \ifodd #1
   \setlength{\leftmargin}{\cslhangindent}
   \setlength{\itemindent}{-1\cslhangindent}
  \fi
  \setlength{\itemsep}{#2\baselineskip}}}
 {\end{list}}
\usepackage{calc}

\usepackage{amsmath, amssymb, amsfonts, mathtools}
\usepackage{siunitx, bm} 

\usepackage{graphicx, xcolor, float, tcolorbox, wrapfig}
\usepackage{caption, subcaption}

\usepackage{booktabs, multirow, multicol, makecell, tabularx, longtable}
\usepackage{threeparttable, threeparttablex, csvsimple}

\usepackage{booktabs}
\usepackage{longtable}
\usepackage{array}
\usepackage{multirow}
\usepackage{wrapfig}
\usepackage{float}
\usepackage{colortbl}
\usepackage{pdflscape}
\usepackage{tabu}
\usepackage{threeparttable}
\usepackage{threeparttablex}
\usepackage[normalem]{ulem}
\usepackage{makecell}
\usepackage{xcolor}

\usepackage{hyperref}
\hypersetup{colorlinks=true, allcolors=blue}

\usepackage{fancyhdr}
\fancypagestyle{plain}{
  \fancyhf{}
  \fancyhead[L,R]{\small\color{red}PREPRINT/WORKING PAPER}
  \fancyhead[C]{\small\color{red}\url{https://krishpn.github.io}}
  \fancyfoot[L,R]{\small\color{red}PREPRINT/WORKING PAPER}
  \fancyfoot[C]{\thepage}
}

\KOMAoption{captions}{tableheading}
\makeatletter
\@ifpackageloaded{caption}{}{\usepackage{caption}}
\AtBeginDocument{%
\ifdefined\contentsname
  \renewcommand*\contentsname{Table of contents}
\else
  \newcommand\contentsname{Table of contents}
\fi
\ifdefined\listfigurename
  \renewcommand*\listfigurename{List of Figures}
\else
  \newcommand\listfigurename{List of Figures}
\fi
\ifdefined\listtablename
  \renewcommand*\listtablename{List of Tables}
\else
  \newcommand\listtablename{List of Tables}
\fi
\ifdefined\figurename
  \renewcommand*\figurename{Figure}
\else
  \newcommand\figurename{Figure}
\fi
\ifdefined\tablename
  \renewcommand*\tablename{Table}
\else
  \newcommand\tablename{Table}
\fi
}
\@ifpackageloaded{float}{}{\usepackage{float}}
\floatstyle{ruled}
\@ifundefined{c@chapter}{\newfloat{codelisting}{h}{lop}}{\newfloat{codelisting}{h}{lop}[chapter]}
\floatname{codelisting}{Listing}

\makeatother
\makeatletter
\@ifpackageloaded{caption}{}{\usepackage{caption}}
\@ifpackageloaded{subcaption}{}{\usepackage{subcaption}}
\makeatother
\usepackage{bookmark}
\IfFileExists{xurl.sty}{\usepackage{xurl}}{} 
\hypersetup{
  pdftitle={Detecting and Explaining Unlawful Insider Trading: A Shapley Value and Causal Forest Approach to Identifying Key Drivers and Causal Relationships},
  pdfauthor={Krishna Neupane; Igor Griva; Robert Axtell; William Kennedy; Jason Kinser},
  colorlinks=true,
  linkcolor={blue},
  filecolor={Maroon},
  citecolor={Blue},
  urlcolor={Blue},
  pdfcreator={LaTeX via pandoc}}

\title{Detecting and Explaining Unlawful Insider Trading: A Shapley
Value and Causal Forest Approach to Identifying Key Drivers and Causal
Relationships}
\author{Krishna Neupane \and Igor Griva \and Robert Axtell \and William
Kennedy \and Jason Kinser}
\date{2025-03-02}
\begin{document}
\maketitle
\begin{abstract}
Corporate insiders trade for various reasons, often possessing Material
Non-Public Information (MNPI). Accurately determining whether a trade
was conducted with MNPI is a challenging task due to its complexity. The
work presented here focuses on two critical aspects, first, accurately
detecting Unlawful Insider Trading (UIT), and second, identifying the
key features explaining accurate classification. It demonstrates how
combining Shapley Values (SHAP) and Causal Forest (CF) can effectively
reveal these explanatory features. The findings reported herein
underscores the importance of causality in effectively identifying,
interpreting, and explaining the UIT. It necessitates considering
alternative scenarios and their potential outcomes. Within a
high-dimensional feature space, the proposed architecture integrates
state-of-the-art techniques. The resulting framework exhibits high
classification accuracy and provides robust feature rankings through
SHAP and causal significance assessments using CF, facilitating the
discovery of unique causal relationships. The analysis demonstrates
statistically significant relationships between the outcome and several
key features, including director status, price-to-book ratio, return,
and market beta. These features significantly influence the likelihood
of the outcome, suggesting potential links between insider trading
behavior and factors such as information asymmetry, valuation risk,
market volatility, and stock performance. This analysis draws attention
to the complexities inherent in financial causality, suggesting that
while initial descriptors may offer intuitive insights, a deeper
examination is required to fully understand their nuanced and often
uncertain impacts. Nevertheless, these findings reaffirm the
architectural flexibility of decision tree models. By incorporating
heterogeneity during tree construction, these models effectively uncover
latent structures within trade, finance, and governance data, thereby
characterizing fraudulent and non-fraudulent behavior while maintaining
reliable results.

\vspace{1em}

\noindent © Authors 2025. All rights reserved. This working paper is
part of on going review for publication. No part of this publication may
be reproduced without prior permission. \url{https://krishpn.github.io}

\vspace{1em}

\noindent \textbf{Keywords:} insider trading, Shapley Values (SHAP),
average treatment effect, causal forest, machine learning, explanation,
interpretability, augmented inverse propensity score, XGBoost,
informational asymmetry, market microstructure

\vspace{1em}

\noindent \textbf{JEL:} G14, K22, C45, C21, G12, D82
\end{abstract}

\setstretch{1}
\section{Introduction}\label{sec-introduction}

Corporate insiders, by virtue of their privileged access to confidential
company information, possess a unique advantage in the market. Despite
stringent regulations, some insiders engage in Unlawful Insider Trading
(UIT) by exploiting material non-public information (MNPI) for personal
gain, breaching fiduciary duties and trust. These unlawful trades often
exhibit complex patterns, blending characteristics associated with
value, growth, momentum, and profitability, making it challenging to
distinguish ``routine'' trading from ``opportunistic'' UIT (Cohen,
Malloy, and Pomorski (2012)).

To address this complexity, this study employs a comprehensive feature
set, including variables related to firm characteristics, market
microstructure, and insider attributes. Among these features, the
Director variable and the Price-to-Book ratio (P/B) emerged as
particularly salient indicators for testing the core legal and economic
theories of insider trading, including misappropriation and the
existence of information asymmetry. Specifically, the Director variable
is examined, as a director's trade provides the clearest indication of
activity stemming from a fiduciary who possesses non-public, material
information (Kallunki, Nilsson, and Hellström (2009), Seyhun (1992),
Seyhun (1986)). Similarly, the Price-to-Book ratio (P/B) is analyzed to
capture the potential profitability of this asymmetric information.
Insiders are economically motivated to trade when they believe the
market has mispriced the firm, and a significant deviation in the P/B
ratio often serves as a proxy for such mispricing, indicating high
potential returns from their superior knowledge (Piotroski (2000),
Lakonishok and Lee (2001)).

Academic research on informed trading, including the activity driven by
directors and P/B-related mispricing, often focuses on two primary,
interacting avenues: information and liquidity.

\subsection{Information-Based
Theories}\label{information-based-theories}

First, researchers using the information-based approach examine how
private information, often held by corporate insiders, influences asset
prices. This perspective draws heavily from the principal-agent problem
Akerlof (1970), which describes the inherent conflict of interest
arising from information asymmetry. In the context of financial markets,
this asymmetry means that insiders typically possess more information
about a company's prospects than outside investors. This informational
advantage allows insiders to potentially profit from trading on this
private knowledge, leading to asset price adjustments that reflect the
incorporation of this information Easley et al. (1996), Kyle (1985). For
example, if insiders know that a company is about to announce
unexpectedly strong earnings, they might buy the stock before the
announcement, driving up the price.

\subsection{Liquidity Risk and
Asymmetry}\label{liquidity-risk-and-asymmetry}

The second primary avenue focuses on liquidity risk and asymmetry,
recognizing that the cost of trading---and the risk of being unable to
trade quickly without affecting the price---is not uniform across
assets. This liquidity dimension is explored through two complementary
lenses: systematic and dynamic risk.

\subsubsection{\texorpdfstring{Systematic Liquidity Risk
(\(\beta_{SMB}\))}{Systematic Liquidity Risk (\textbackslash beta\_\{SMB\})}}\label{systematic-liquidity-risk-beta_smb}

The \textbf{systematic component} of this risk is often proxied using
factors linked to structural illiquidity. Specifically, the SMB Beta
(\(\beta_{SMB}\)) from the Fama and French (1993) model serves as a
common proxy for exposure to this risk. Small-cap stocks, which are
measured by the Small Minus Big (SMB) factor, are historically less
liquid than large-cap stocks Banz (1981). Therefore, a security's
\(\beta_{SMB}\) reflects its sensitivity to the size premium, which is
widely theorized to compensate investors for bearing the higher,
systematic risk of illiquidity and the associated information asymmetry
inherent in smaller, less frequently traded firms.

\subsubsection{Market Liquidity
Dynamics}\label{market-liquidity-dynamics}

A complementary body of work emphasizes the role of market liquidity
dynamics in asset pricing. Market liquidity refers to the ease with
which an asset can be bought or sold without significantly impacting its
price. This approach posits that the dynamic interplay between the
demand for and supply of liquidity is a crucial determinant of asset
prices (\textbf{harris1998trading?}),
(\textbf{pagano1996transparency?}), (\textbf{admati1988theory?}). For
instance, if there is high demand for a particular stock but limited
supply, the price is likely to increase. Conversely, if many investors
are trying to sell a stock but there are few buyers, the price might
fall.

It is important to note that these perspectives interact: information
asymmetry, driven by informed trades such as those by directors
exploiting P/B mispricing, can also play a direct role in liquidity. If
outside investors suspect that insiders have private information, they
may be less willing to trade, reducing liquidity and potentially
amplifying price movements. Therefore, while the information and
liquidity perspectives offer distinct explanations, they are not
necessarily mutually exclusive and can interact to influence asset
prices.

\subsection{Legal and Regulatory
Frameworks}\label{legal-and-regulatory-frameworks}

Alternatively, legal scholars have offered several theories to explain
insider trading, including the ``equal access'' theory (Bines (1976)),
which emphasizes the unfair advantage gained from possessing non-public
information. Other theories focus on the misappropriation of such
information, particularly when it is obtained through one's position
within a company (Gervasi (2023)). A further perspective views ITas a
form of fraud, involving affirmative misrepresentation to deceive
investors (Bondi and Lofchie (2011)). However, these diverse theories
have not converged on a unified understanding of insider trading.
Consequently, proposed solutions range from complete deregulation,
arguing that allowing Insider Trading (IT) incentivizes innovation
(Bainbridge (2022), Manne (1966)), to strict prohibitions aimed at
preventing managerial overreach and unlawful profits (Gangopadhyay and
Yook (2022)). This lack of consensus among legal scholars contributes to
the often-vague characterization of UIT as simply a result of
``mispricing'' (Huang and Tung (2021)). This vagueness stems, in part,
from the difficulty in definitively identifying which theory, if any,
fully explains the phenomenon.

\subsection{Limitations of Traditional
Methods}\label{limitations-of-traditional-methods}

While these theoretical frameworks offer valuable insights, they often
rely on traditional statistical methods that, despite their
contributions, suffer from significant limitations in today's data-rich
environment (Mayo and Hand (2022), Mazzarisi et al. (2022), West and
Bhattacharya (2016)). These manual techniques often lack the flexibility
to adapt to complex datasets, can lead to mis-specified models that fail
to capture the true relationships between variables, and struggle to
adequately represent the interactive effects between multiple factors.
Furthermore, the lack of standardized procedures can make replicating
studies challenging, hindering the accumulation of robust findings. A
prime example of these practical limitations is the common practice of
estimating relationships via regression coefficients using a limited set
of variables. This approach, exemplified by methods like Auto Regressive
Moving Average (ARIMA) modeling often struggles with scalability as the
volume of available data increases. As datasets grow larger and more
complex, these traditional techniques become increasingly impractical
for uncovering the intricate dynamics of the phenomena under study
(Cerniglia and Fabozzi (2020), Hand (2009)).

\subsection{Computational Methodology}\label{computational-methodology}

In today's data-driven landscape, advanced supervised learning
techniques have demonstrated superior performance in detecting unusual
patterns and classifying UIT, as illustrated in studies like Deng et al.
(2021) and Deng et al. (2019). A key advantage of these supervised
classifiers is their inherent model interpretability (Eggensperger et
al. (2018)). They provide insights into the relative importance of
different features in distinguishing between lawful and unlawful
transactions. For example, in Neupane and Griva (2024b) ranks features
to enhance explainability. However, a limitation of feature rankings
derived directly from these models is that they are based on the
training data and may not fully account for correlations between
features. To address this, this paper goes a step further by removing
these correlations and re-ranking features in the test data using
SHapley Additive exPlanations (SHAP) values introduced by Shapley et al.
(1953). SHAP values estimate the marginal contribution of each feature
across all possible feature combinations (S. M. Lundberg, Erion, and Lee
(2019)), offering a more nuanced understanding of feature importance.
Even with highly correlated and heterogeneous trade and financial data,
SHAP value ranking serves as a valuable intermediate step in identifying
the leading factors associated with UIT.

To further validate the credibility and reliability of the top-ranked
predictors identified through SHAP values, this paper employs Causal
Forest (CF) to uncover the underlying causal relationships between these
features. CF, initially proposed by Athey and Imbens (2016), utilizes a
bootstrap aggregation approach similar to Random Forest (RF). It is
specifically designed to address ``what if'' or ``what would''
questions, leveraging the concept of ``potential outcomes,'' where each
outcome corresponds to a different level of treatment manipulation
(Imbens and Rubin (2015)). Over the past decade, causal inference
methods have seen tremendous growth in applications across diverse
fields, including accounting (Gow, Larcker, and Reiss (2016)), asset
pricing (Hiemstra and Kramer (1997), Cready, Kumas, and Subasi (2014)),
education (Athey and Wager (2019)), agriculture (Deines, Wang, and
Lobell (2019)), corporate investment (Gulen, Jens, and Page (2022)), and
spatial data analysis (Credit and Lehnert (2023)). This study extends
the feature importance analysis of UIT previously conducted by Neupane
and Griva (2024b) using XGBoost. Building upon this foundation, we focus
on interpreting the causal significance of the identified features.
While a complementary analysis by a subset of the authors (Neupane and
Griva (2024a)) also demonstrated insightful feature rankings using RF,
the current investigation prioritizes XGBoost due to its widespread
adoption and performance.

\subsection{Contributions}\label{contributions}

This research makes several key contributions to the study of UIT.
First, it pioneers the application of a combined
classification-causality approach. This innovative methodology
simplifies the complex task of uncovering the root causes of UIT by
leveraging latent regularities and underlying data attributes that
reflect the choices made by insiders. By combining the predictive power
of classification techniques with the explanatory power of causal
inference, this approach moves beyond simply identifying suspicious
transactions to understanding why they occur. Second, this research
tackles several limitations inherent in traditional methods. It
addresses the challenges posed by hand-engineered features, omitted
variables, interdependencies between variables, and the high
dimensionality of financial data. These issues often hinder the
understanding, explanation, and replication financial anomalies (see
Hou, Xue, and Zhang (2020) for details). By employing advanced machine
learning and causal inference techniques, this study offers a more
robust and comprehensive approach. Third, the research aims to uncover
hidden patterns of IT that may have been overlooked by traditional
econometric methods, particularly those susceptible to ``p-hacking'' --
the manipulation of data or statistical analyses to achieve
statistically significant results. By using a more rigorous and
data-driven approach, this study seeks to provide a more reliable
characterization of IT behavior.

The paper proceeds as follows: Section~\ref{sec-method-proposed-method}
details the proposed methodology, including the SHAP value approach for
feature ranking. Section~\ref{sec-analysis-experimental-setup} describes
the experimental setup used in the study.
Section~\ref{sec-analysis-results-data} presents the classification
results, including metrics from the confusion matrix. It then presents
and discusses the relative feature rankings based on SHAP values, both
before and after correlation removal, and the results from the CF
analysis. Finally, Section~\ref{sec-conclusions-future} summarizes the
key findings and outlines potential avenues for future research.

\section{Proposed Method}\label{sec-method-proposed-method}

The overall approach for this analysis integrates three distinct machine
learning methodologies to ensure a robust and interpretable assessment
of the data: predictive modeling (XGBoost), model interpretation (SHAP
values), and causal inference (CF).

In the realm of supervised learning, tree-based ensemble methods like RF
and XGBoost have consistently outperformed other techniques, including
deep learning models, particularly when dealing with tabular data
(Shwartz-Ziv and Armon (2022), Borisov et al. (2022), Gorishniy et al.
(2021), Grinsztajn, Oyallon, and Varoquaux (2022)). Their effectiveness
is further underscored by their increasing prevalence in empirical
economics, often serving as a crucial intermediate step in the
analytical process (Athey (2019)).

Building on the established success of these powerful ensemble methods,
this paper introduces the implementation of CF, a sophisticated
extension that goes beyond prediction to explore the ``why'' behind
observed patterns. CF delves into the data to uncover irregularities
stemming from underlying causal relationships. This section details the
CF methodology, while also briefly introducing other relevant methods
employed in the study.

\subsection{extreme Gradient Boosting (Predictive
Modeling)}\label{sec-method-gradient-boosting}

The XGBoost algorithm will first be employed for feature selection and
predictive modeling due to its superior performance in structured data
settings. This method is utilized to identify the most salient features
impacting the target variable.

XGBoost was proposed by Chen and Guestrin (2016) as an extension of
generalized gradient boosting aimed at handling large datasets. The
underlying greedy learning method supports parallelization that
iteratively updates the weights of weakly created trees (base learners)
and uses a boosting mechanism to learn from data attributes (Friedman,
Hastie, and Tibshirani (2000)). The repeated re-weighting refines and
improves accuracy by learning from previously unexamined data
attributes. As prediction mistakes are encountered, each prediction is
grouped and fed into the ensemble for weighted voting (Weighted Majority
Algorithm). Pattern detection continues until no further new patterns
are detected by the base learners. In technical terms, the construction
of the new base learners is maximally correlated with the negative
gradient of the loss function. For a detailed technical description,
readers are invited to reference Neupane and Griva (2024b). For the
purpose of this study, the XGBoost model is trained using a 5-fold
cross-validation scheme and early stopping to prevent overfitting.

\subsection{Feature Importance and Model Interpretation using
SHAP}\label{sec-method-feature-importance}

Following the training of the XGBoost model, SHAP values are utilized to
provide crucial model interpretability. SHAP is applied to attribute the
prediction of the XGBoost model to each input feature individually,
calculating the marginal contribution of each feature to the final
prediction. This process ensures transparency and allows for a clearer
understanding of the model's decision-making process.

Ensemble methods, during training, extract, compare, and rank
significant features based on impurity scores (Xu et al. (2014), Duchi
et al. (2008), Schölkopf et al. (2001)). This ranking enhances model
interpretability, explainability, and predictive accuracy in downstream
applications (Qian et al. (2022), Xu et al. (2014), Guyon et al. (2010),
Genuer, Poggi, and Tuleau-Malot (2010), Strobl et al. (2007)). Impurity
scores, such as Gini impurity or entropy, measure the probability of
misclassifying a randomly chosen data point if labels were assigned
randomly based on class distribution (Nembrini, König, and Wright
(2018), Breiman (2001)). However, feature ranking based solely on
impurity scores suffers from limitations, namely that it relies solely
on training data, potentially leading to overfitting and biased feature
importance assessments. Secondly, in circumstances with correlated
features, the ranking can be distorted, making it difficult to
accurately assess the individual contributions of each feature (Shalit
(2020), Roth and Verrecchia (1979)).

To address these shortcomings, SHAP is employed, a method rooted in
cooperative game theory. SHAP treats each feature as a player---these
players collaborate and compete to form coalitions, aiming to maximize
collective gains (expected marginal benefits) without external
enforcement. The ``payoff'' represents the difference in model output
when the feature is included versus excluded from a coalition of other
features. Hence, higher payoffs indicate greater feature importance.
SHAP has been successfully adopted across a wide array of domains, for
example, to study cost allocation in companies (Lemaire (1984)),
investigating valuation due to corporate voting (Zingales (1992)), wage
bargaining between firm and the multiple employees
((\textbf{brugemann2019intra?})), financial fraud detection (Lin and Gao
(2022)), and measuring the attribution of risks in the banking system
(Tarashev, Tsatsaronis, and Borio (2016)) and so on. Importantly, SHAP
satisfies four key axioms (Wang et al. (2024)):

\begin{itemize}

    \item Efficiency: Fair and complete distribution of total value among all features.
    \item Symmetry: Equal contributions from features result in equal rewards.
    \item Dummy: Features with no contribution receive zero payoff.
    \item Additivity: The SHAP of the sum of two value functions equals the sum of their individual SHAP.
\end{itemize}

Formally, in a fair distribution model \(f\) with \(d\) input features,
features cooperate to form combinations by producing classification
values. For any set of features \(S\), \(f(S)\) is the classification
value based on that set. Shapley's formulation considers the incremental
value brought to a set \(S\) if a feature \(x_j \notin S\) were to join
it. This incremental value (marginal contribution) of feature \(x_j\)
given that the features in \(S\) are already present is represented by
Equation \ref{eq:causaility marginalcontribution} (Mase, Owen, and
Seiler (2022), Kamath et al. (2021), S. Lundberg and Lee (2017)).

\begin{equation}
    \begin{aligned}
        f(x_j|S)=f(S \cup\{x_j \})-f(S),
        \label{eq:causaility marginalcontribution} 
        \end{aligned}
    \end{equation}

where, \(x_j\) represents the individual feature whose contribution is
being assessed; \(f\) denotes the model's prediction output based on the
available features; \(S\) represents the set of features currently
included in the model, excluding feature \(x_j\); \(S \cup \{x_j\}\)
represents the set of features \(S\) with feature \(x_j\) added to it;
\(f(S \cup \{x_j\}))\) represents the model's prediction when both \(S\)
and \(x_j\) are considered; and~ \(f(S)\) represents the model's
prediction when only the subset \(S\) is considered.

Equation \ref{eq:causaility marginalcontribution} evaluates how much the
model's predictive power improves when feature \(x_j\) is added, given
that other features in \(S\) are already included. The Shapley value
\(\phi_j\) for the \(j^{th}\) feature is estimated with Equation
\ref{eq:causaility shapleyMainEquation}:

\begin{equation}
    \begin{aligned}
           \phi_j(f)=\sum_{S\in P( X \backslash  \{x_j\})} \frac{|S|!(d-|S|-1)!}{d!} \left[ f \left( S \cup \{x_j \}\right) -f(S) \right]
            \label{eq:causaility shapleyMainEquation} 
        \end{aligned}
    \end{equation}~

where, in a given set of features (\(X\)) of \(d\) dimensions,
\(\phi_j(f)\) represents the SHAP value for feature \(x_j\); \(S\)
represents a subset of features that excludes \(x_j\);
\(P(X \backslash  \{x_j\})\) denotes all possible combinations of the
other features without \(x_j\); (\(|S|!\)) accounts for the ways to add
(\(x_j\)) after all features in (\(S\)) have been considered;
(\((d-|S|-1)!\)) accounts for the ways to add (\(x_j\)) before the
remaining features not in (\(S\)); (\(d!\)) normalizes this weight,
ensuring all weights sum to 1 across all permutations of features;
(\(f(S)\)) is the model's predictive value with only the features in
subset \(S\); and \(f(S \cup \{x_j\})\) represents the model's
prediction when adding feature \(x_j\) to the subset \(S\). The SHAP
value (\(\phi_j(f)\)) represents the average marginal contribution of
\(x_j\) across all possible orders of adding features to the model. It
quantifies the importance of \(x_j\) in the prediction \(f\) by
considering how much its inclusion changes the prediction, on average,
across all possible scenarios of feature inclusion.

\subsection{Multicollinearity
Diagnostics}\label{sec-method-causal-forests}

Following feature selection, multicollinearity among the final set of
predictors was quantified using the Variance Inflation Factor (VIF)
analysis. An iterative filtering process was applied to the initial set
of 110 features, repeatedly removing the feature with the highest VIF
score until all remaining features were below the conservative threshold
of 10.0.

This diagnostic step is critical to ensure the stability and reliability
of the treatment effect estimates generated by the CF. The VIF scores
for the key features utilized in the final model were confirmed to be in
the low range: Price-to-Book showed a VIF of 1.58, and Market Beta
(\(\beta\)) showed a VIF of 1.82. The highest VIF observed across the
final set of variables was 9.41. These results confirm that the model
inputs are sufficiently independent to ensure robust causal inference.

\subsection{Heterogeneity Analysis (Hierarchical
Clustering)}\label{sec-method-Hierarchical-Clustering}

To empirically derive stable subgroups for the subsequent causal
analysis, and to provide a diagnostic visualization of feature
dependencies, Hierarchical Clustering was applied to the transaction
data. This method uses the final, VIF-filtered set of features to group
observations based on similar financial and trading characteristics. The
resulting dendrogram was used as a primary tool to visualize clusters of
highly correlated features, guiding the iterative VIF-based feature
selection process. Unlike pre-specified or a priori subgroups,
hierarchical clustering allows for the data-driven discovery of
naturally occurring clusters of transactions that exhibit shared
attributes and behavior profiles. The resulting cluster assignments are
utilized to stratify the treatment effect estimation in the Causal
Forest model. This ensures that the Conditional Average Treatment
Effects (CATE) are estimated on genuinely distinct sub-populations,
thereby strengthening the robustness and granularity of the causal
conclusions.

\subsection{Causal Forests}\label{sec-method-causal-forests}

Finally, to address potential confounding variables and estimate
treatment effects, a CF model is implemented. This method is used to
estimate the heterogeneous treatment effects of the key variable of
interest across various subgroups within the dataset. The CF provides an
estimate of the conditional average treatment effect (CATE), allowing
for robust causal conclusions in the absence of a controlled
experimental design.

CF, a non-parametric generalization of RF, was developed by Athey and
co-authors. Their methodological contributions are detailed in several
publications, including Nie and Wager (2021), Athey and Wager (2021),
Athey, Tibshirani, and Wager (2019), Athey and Wager (2019), Athey
(2019), Wager and Athey (2018), and Athey and Imbens (2016). In a sense,
CF is an ``improvement'' to understand causal effects rather than a
refinement of prediction accuracy, which marks a fundamental shift from
traditional machine learning. The method addresses the fundamental
question of causality: ``what if'' or ``what would'' happen under
different conditions or interventions (Imbens and Rubin (2015)). The
notion of causality addresses questions pertaining to ``potential
outcomes'' and ``treatment'' by comparing between ``actual'' to the
``counterfactuals'' (Cunningham (2021)). However, there is an inherent
inability to observe both potential outcomes (factuals and
counterfactuals) simultaneously: the outcome under treatment and the
outcome under control. Such inability to observe the ``counterfactual''
outcome is known as the ``fundamental problem of the causal inference''
(Gelman (2011), Pearl (2010), Angrist and Imbens (1995), Holland
(1986)).

CF estimates causal effects in complex, heterogeneous, and non-linear
data through the estimation of Conditional Average Treatment Effects
(CATE) (Athey and Wager (2019), Athey, Tibshirani, and Wager (2019)).
CATE provides a deepened understanding of how treatment effects vary
across each transaction level. The uniqueness of the CATE is that it
improves upon its predecessor, the Average Treatment Effect (ATE). ATE
represents the average difference in outcomes for the entire population
under treatment and control conditions (Jawadekar et al. (2023), Athey
and Wager (2019), Athey, Tibshirani, and Wager (2019)). However, relying
solely on ATE can be misleading and spurious, as it may obscure
important subgroup-level variations in treatment effects (Cook, Gebski,
and Keech (2004), Assmann et al. (2000)), particularly if sub-groups are
assigned. Even though pre-specification of sub-groups can provide some
insights, it ``can make it difficult to discover strong but unexpected
treatment effect heterogeneity'' (Wager and Athey (2018)). The
assignment to the sub-group may also raise ethical questions. The
Average Treatment Effect (ATE) across the entire population represents
the difference in potential outcomes (\(Y_i\)) attributable to the
treatment effects (\(W_i\)), as defined in Equation
\ref{eq-ate-estimation}. Note that, in observational studies, the
inability to observe ``counter-factuals'' is a persistent challenge.

    \begin{equation}
        \begin{aligned}
            \text{ATE} = \frac{1}{N} \sum_i [ \mathbb{E}[Y_i^{w=1}] -  \mathbb{E}[Y_i^{w=0}]], 
            \label{eq-ate-estimation} 
        \end{aligned}
    \end{equation} 
where, \(\mathbb{E}[Y_i^{w=1}]\) and \(\mathbb{E}[Y_i^{w=o}]\) are respectively expected value of the outcome if transaction \(i\) receives or does not receive the treatment normalized by \(\frac{1}{N}\), the sum over all transactions \(N\).

To determine the treatment effect at the individual transaction level,
CATE is estimated (see Equation \ref{eq-cate-estimation}), an
improvement over ATE that accounts for the feature vector (\(x \in X\)).

    \begin{equation}
        \begin{aligned}
            \text{CATE} (\tau(x)) = \mathbb{E}[Y_i^{w=1} - Y_i^{w=0} \mid X_i = x], 
            \label{eq-cate-estimation} 
        \end{aligned}
    \end{equation} 
where, \(\tau(x)\) estimates the treatment effect between potential outcomes of the expectation of \(Y_i^{w=1}\) when treated and \(Y_i^{w=0}\) as represented by \(\mathbb{E}[Y_i^{w=1} - Y_i^{w=0} \mid X = x]\) for transaction \(i\), given that their feature \(X_i\) are equal to \(x\).

In summary, the goal of estimating \(\tau(x)\) is to create leaves where
the treatment effect is different between leaves (heterogeneous) but
similar within each leaf (homogeneous) within single tree. This
hierarchy of node-leaf constructs an individual causal tree. Collection
of single trees are ensembled to produce CF. During forest construction,
CATE estimates at each tree level is averaged (see Equation
\ref{eq-froest-construction}). The treatment effect heterogeneity by
segmenting the transactions into various sub-groups based on feature in
consideration both at \(node\) and \(leaf\) level. At the \(node\)
level, during the split subgroups are created based on the close feature
values, thus creating differences between the nodes. Since input \(x\)
are different, the estimated \(\tau(x)\) is which when averaged are
different. A \(leaf\) is the terminal \(node\) where no further split
can be made. The \(\tau(x)\) at \(leaf\) level estimates the treatment
effect for all transaction that fall into leaf based on feature profile
(\(x\)).

    \begin{equation}
        \begin{aligned}
            \text{CATE}_\text{forest}(X_i) = \frac{1}{B} \sum_{b=1}^B \text{CATE}_{\text{tree}_b}(X_i), 
            \label{eq-froest-construction} 
        \end{aligned}
     \end{equation}
     where, $B$ is the number of trees in the forest, and CATE $\text{CATE}_{\text{tree}_b}(X_i)$ is the CATE estimate for each transaction $(X_i)$ from the tree $b$. Each tree in the forest, indexed by $b$ from 1 to $B$, provides its own estimate of the CATE for each transaction with features $X_i$. The final The CATE estimate is average across all $B$ trees. 

Athey and Imbens (2016) implements Augmented Inverse Probability
Weighting (AIPW, see Equation \ref{eq-aipw-equation}) for estimating
CATE. This choice is motivated methodological synergy that the method
effectively handles complex data and provides straightforward analysis
of heterogeneous treatment effects. It also addresses confounding by
weighting observations based on their propensity scores, adjusting for
the probability of receiving the observed treatment. The AIPW estimator
thus involves two fundamental steps: first, it estimates the probability
of the treatment assignment conditional on feature (observed). The
outcome of this step is then weighted by the propensity score to produce
weighted average. Propensity score is the probability that transaction
would be assigned to a particular treatment group based on the observed
feature (Rosenbaum (2023), (\textbf{rosenbaum1983central?})).

    \begin{equation}
        \begin{aligned}
            \hat{\tau}_{AIPW _i} =\frac{W_i(Y_i-\hat{\mu}_1(X_i))}{\hat{e}(X_i)} + \frac{(1-W_i)(Y_i-\hat{\mu}_0(X_i))}{1-\hat{e}(X_i)}+(\hat{\mu}_1(X_i)-\hat{\mu}_0(X_i)), 
            \label{eq-aipw-equation} 
        \end{aligned}
     \end{equation}
    where, for transaction $i$, $Y_i$ is the observed outcome; $W_i$ is the treatment indicator; $\hat{e}(X_i)$ is the propensity score, the estimated probability of treatment given $X_i$; and $\hat{\mu}_1(X_i)$ and $\hat{\mu}_0 (X_i)$ are the predicted outcomes under treatment and control given features $X_i$, respectively.

In Equation \ref{eq-aipw-equation}, the difference between
\(\hat{\mu}_1(X_i)\) - \(\hat{\mu}_0(X_i)\) is an adjustment term that
represents the estimated treatment effect based on the outcome models.
This term helps to reduce the variance of the AIPW estimator by
leveraging information from the outcome models. The first two terms of
Equation \ref{eq-aipw-equation} are crucial for balancing the treated
and control groups by contributing weights to the observed outcomes.

When \(W_i = 1\) (transaction \(i\) is treated), the first term,
\(\frac{W_i(Y_i - \hat{\mu}_1(X_i))}{\hat{e}(X_i)}\), gives more weight
to observations where the propensity score \(\hat{e}(X_i)\) is lower.
This means transactions with features \(X_i\) that made treatment less
likely receive higher weights, aiming to balance the treated group with
the untreated group in terms of feature distribution. The second term,
\(\frac{(1-W_i)(Y_i - \hat{\mu}_0(X_i))}{1-\hat{e}(X_i)}\), drops out
because \((1-W_i)\) becomes zero, meaning it does not contribute to the
estimate for treated transactions. If \(\hat{e}(X_i) < 1\), then
\((1 - \hat{e}(X_i))\) will be positive, adjusting the estimate based on
how much less likely treatment was than a perfect prediction.
\(\hat{e}(X_i)\) cannot be greater than 1 as it represents the
propensity score, which is the probability of receiving treatment given
features \(X_i\). If \(\hat{e}(X_i) = 1\), the denominator becomes zero,
highlighting the importance of overlap.

Similarly, when \(W_i = 0\) (transaction \(i\) is untreated), the first
term drops out because \(W_i\) becomes zero, meaning untreated
transactions do not directly contribute to its estimate. The second
term, \(\frac{(1-W_i)(Y_i - \hat{\mu}_0(X_i))}{1-\hat{e}(X_i)}\),
provides weights to the observed outcome by the inverse of the
probability of \textit{not} being treated, \((1 - \hat{e}(X_i))\). This
gives more weight to observations where treatment was more likely
(higher \(\hat{e}(X_i)\)), thus balancing the untreated group with the
treated group. If \(\hat{e}(X_i) > 0\), then \(\hat{e}(X_i)\) will be
positive, adjusting the estimate based on how much more likely treatment
was. Finally, if \(\hat{e}(X_i) = 0\), it means there is a zero
probability for someone with those features \(X_i\) to receive
treatment, which indicates a lack of overlap.

Building upon the foundation of CATE, CF proceeds as follows:

\begin{enumerate}
    \item  \textbf{Bootstrap Sampling}: Each tree in the forest is constructed using a bootstrap sample drawn with replacement from the original dataset. This introduces randomness, as each tree observes a slightly different version of the data, capturing data variability and improving the model's overall robustness. 
    \item \textbf{Subsampling}: To further reduce variance and prevent overfitting, only a portion of the bootstrap sample is used to build each tree. This limits the impact of any idiosyncrasies within the full dataset, leading to better generalization performance. 
    \item \textbf{Honesty Split}: To avoid overfitting, the subsample is divided into two parts: a ``splitting'' set used to determine the tree structure (node splits) and an ``estimation'' set used to estimate treatment effects or other parameters within each leaf node. Therefore, the ``honesty'' procedure prevents from overfitting to the same data used for both tree construction and parameter estimation.
    \item \textbf{Tree Construction}: CF diverge fundamentally from the RF in their tree-building process. RF prioritizes splits that minimize within-node outcome variance, akin to criteria like information gain.CF, however, focuses on maximizing the difference in treatment effects between the resulting child nodes. This "variance improvement" in CF is quantified by the squared difference in treatment effects: $\Delta \tau^2 = (\tau_{\text{left}} - \tau_{\text{right}})^2$, where  $\tau_{\text{left}}$ \text{ and } $\tau_{\text{right}}$ \text{represent the treatment effects.} represent the treatment effects in the left and right child nodes, respectively. To identify the optimal split, CF exhaustively evaluates all possible split points and selects the one that yields the highest squared difference in treatment effects, pinpointing where the treatment effect diverges most significantly.
    \item \textbf{Aggregation}: Final predictions are obtained by averaging the CATE estimates from all individual trees in the forest. This averaging process reduces variance and aims to produce a more robust estimate of the treatment effect for each individual (see Equation \ref{eq-froest-construction}). 
\end{enumerate}

Hence, the average causal effect (expected utility) on the potential
outcomes from the deployment of treatments subject to constraints and
functional forms is measured (Jacob (2021), Athey, Tibshirani, and Wager
(2019), Rubin (1980)).

\subsection{Performance Measure}\label{sec-method-performance-measure}

The performance of binary supervised classification models is commonly
assessed using a 2 \(\times\) 2 confusion matrix, schematically
illustrated in Table \ref{tab:confusionMatrixSchematic}. This matrix
juxtaposes actual and predicted class labels along its rows and columns,
respectively (Hastie, Tibshirani, and Friedman (2009)).

\begin{table}
    \centering
    \caption{Confusion Matrix Illustrating Predicted vs. Actual Outcomes for Lawful/Unlawful Classes.}
    \label{tab:confusionMatrixSchematic}
    \begin{tabular}{lcc}
        \toprule
        \multicolumn{1}{c}{} & \multicolumn{2}{c}{Predicted Labels (PP+PN)} \\
        \cmidrule(lr){2-3}
        Tot. Pop. (P+N) & Lawful(P) & Unlawful(N) \\
        \midrule
        Lawful(+) & True Lawful (TP) & False Unlawful (FN) \\
        Unlawful(-) & False Lawful (FP) & True Unlawful (TN) \\
        \bottomrule
    \end{tabular}
    \label{tab:confusionMatrixSchematic}
\end{table}

\subsection{Measure of Causal
Significance}\label{sec-method-level-of-signfinace}

To assess causal significance, the analysis employs the \(p\)-value. A
significance threshold (\(\alpha\)) is pre-defined, and results are
interpreted based on whether the \(p\)-value is less than or equal to
\(\alpha\). The \(p\)-value represents the probability of observing the
current data (or more extreme data) if the null hypothesis were true. In
essence, it quantifies the evidence against the null hypothesis
((\textbf{schervish1996p?})). However, it does not directly indicate the
truth or falsity of the null hypothesis. Instead, it provides evidence
regarding the likelihood of observing the current data under the
assumption that the null hypothesis is true. For the purpose of this
analysis, the null hypothesis is that the top \(k\) features (the
treatments) do not explain the underlying causality. The null hypothesis
then is tested the alternative hypothesis that top \(k\) features,
acting as proxies, do indeed explain the UIT.

\section{Experimental Setup}\label{sec-analysis-experimental-setup}

Data sources, pre-processing, and classification model settings are
replicated from Neupane and Griva (2024b). A key innovation of this work
lies in the inclusion of causality analyses, extending beyond the scope
of previous studies. The primary dataset comprises Statement of Changes
in Beneficial Ownership (Form 4) filings obtained from the Securities
and Exchange Commission's (SEC) Electronic Data Gathering, Analysis, and
Retrieval (EDGAR) system. These filings, submitted by individual
insiders within 48 hours of a transaction, are supplemented with trade
and financial data sourced from the CRSP and Compustat-CapitalIQ
databases, merged based on unique identifiers (cik, personid, companyid,
govkey, permano). The resulting dataset encompasses 110 features
characterizing ownership, corporate governance, profitability, financial
performance, risk, and market returns.

In order to label transactions as lawful or unlawful, the methodology
leverage publicly available SEC court complaints. The owners of Form 4
are matched against defendant names in these complaints using the
Levenshtein distance algorithm. Matches with scores exceeding 85 percent
are considered valid. This process identifies 1992 unlawful
transactions. Data preprocessing includes one-hot encoding categorical
variables (Acquisition, Disposition, IsDirector, IsOfficer, and
IsTenPercentOwner) and z-score normalization of numerical features to
achieve a standard normal distribution (mean (\(\mu\)) = 0, standard
deviation (\(\sigma\)) = 1). The balanced dataset comprise
\(0.5 \colon 0.5\) ratio of the lawful to unlawful respectively for
\(320\) and \(3984\) transactions. Following the classification, the
features are ranked determined by SHAP in the test dataset. Ranking are
based on two stages before and after removal of correlation between
features. To address collinearity, hierarchical clustering based on
Spearman rank-order correlation is implemented. A threshold (based on
distance) is defined, and a single feature is selected from each
cluster, resulting in a refined feature ranking.

As a part of the post-hoc analysis, the research investigates the causal
significance of the top \(k\) ranked features. This involves re-ranking
features based on their ability to explain the underlying causality of
UIT. The experiments are conduced with a maximum of 1000 trees in the
forest (number of estimators), the maximum depth of a tree is set at 10;
an honest fraction of ratio 08 : 02 is maintained. The experiments are
conducted using scikit-learn, xgboost, and econml libraries. Model
performance is evaluated using metrics derived from the confusion matrix
and \(p\) values.

\section{Analysis and Results}\label{sec-analysis-results-data}

\begin{table}

\caption{\label{tbl-transactions-all-3984-Count}Illustrative distribution of the unlawful and randomly selected lawful transactions with (0.5:0.5) split ratio. Sub-table in the right-hand side is the subset of randomly selected subset of the left-hand side that matches transactions counts from Deng et. al, (2019). The example is replicated from Neupane and Griva ((2024)).}

\centering{

    \centering
    \begin{minipage}{\linewidth}
    \begin{subtable}{0.45\textwidth}
        \centering
        \begin{tabular}{l | l | l|l}
        Label & Sell & Purchases & Total\\
        \hline \hline
        Lawful & 405 & 1587 & 1992\\
        Unlawful & 318 & 1674 & 1992\\
       \end{tabular}
       
    \end{subtable}
    \hfill
    \begin{subtable}[h]{0.45\textwidth}
        \centering
        \begin{tabular}{l | l | l|l}
        Label & Sell & Purchases & Total\\
        \hline \hline
        Lawful &27 & 133 & 160\\
        Unlawful & 26 & 134 & 160\\
        \end{tabular}
        \label{tbl-transactions-selected-320-Count}
     \end{subtable}
     \end{minipage}
     
    \label{tbl-transactionsCount}

}

\end{table}%

The Section~\ref{sec-analysis-results-data} presents and interprets the
results of the analysis. Table \ref{tbl-transactionsCount} includes
sub-tables adapted from Neupane and Griva (2024b) displaying the
frequency of transactions categorized as lawful and unlawful (split
ratio of 0.5:0.5). To maintain consistency with prior research, a
smaller subset of data is presented alongside the full dataset. Analysis
of the sub-tables reveals a consistent pattern: purchase transactions
are more frequent than sales transactions. This observation aligns with
expectations, given that executive compensation structures often include
restricted stock options and performance-based bonuses, incentivizing
executives to acquire rather than sell company stock (Roulstone (2003)).

\subsection{Results of Classification of IT
Transactions}\label{sec-analysis-Results-Classification-Transactions}

    \begin{ThreePartTable}
        \begin{TableNotes}[para]
            \item[\ddag] Classical Method
            \item[\S] Genetic Algorithm
            \item[\P] Non-Dominated sorting Genetic Algorithm II
        \end{TableNotes}

\begin{longtable}{lrrrrrr}

\caption{\label{tbl-rfComparativeConfusionMatrixBenchMarkMethods}Performance of various metrics according to the benchmark method to identify and detect UIT. Source: Deng et al. 2019}

\tabularnewline

            \toprule
            & ANN & SVM & Adaboost & \multicolumn{3}{c}{XGBoost} \\
            \cmidrule(lr){5-7}
            & & & & Classic\tnote{\ddag} & GA\tnote{\S} & NSGA II\tnote{\P} \\
            \midrule
            ACC & 69.57 & 75.33 & 74.75 & 77.88 & 81.77 & 84.99 \\
            FNR & 19.21 & 21.42 & 26.62 & 22.70 & 16.43 & 13.47 \\
            FPR & 34.07 & 27.75 & 24.42 & 21.56 & 20.10 & 16.31 \\
            PRE & NA & NA & NA & 78.94 & NA & NA \\
            TNR & 65.93 & 72.75 & 75.58 & 78.44 & 79.90 & 83.69 \\
            TPR & 80.79 & 78.58 & 73.38 & 77.30 & 83.57 & 86.53 \\
            \bottomrule
            \insertTableNotes
            \vspace{0.5cm} 

\end{longtable}

    \end{ThreePartTable}

In Section~\ref{sec-analysis-Results-Classification-Transactions}, a
comparative analysis of classification performance metrics, drawing from
prior research. The benchmark model results, sourced from Deng et al.
(2019) and summarized in Table
\ref{tbl-rfComparativeConfusionMatrixBenchMarkMethods}, highlight the
stronger performance of the NSGAII-integrated XGBoost model. This model
demonstrated an overall ACC of 84.99 percent, with a TPR of 86.53
percent for lawful transactions and a TNR of 83.69\% for unlawful
transactions. The FNR and FPR were 13.47 percent and 16.31 percent,
respectively. These results are particularly relevant when considered
alongside the feature importance analysis shown in SHAP plot (see
Section~\ref{sec-analysis-interpretation-and-ranking-covariates} and
Section~\ref{sec-analysis-interpretation-and-ranking-features-after-removal-collinearity-Shapley-based}
for detailed discussion).

    \begin{ThreePartTable}
        \begin{TableNotes}[para]
            \item[\ddag] Subset of 320 random selections from 3984 transactions matching the count of the previous study
        \end{TableNotes}

\begin{longtable}{lrrrr}

\caption{\label{tbl-standaloneRFxGBoost}Comparative mean performance measures derived from 5-fold cross-validation (100 repetitions), contrasting a full transaction dataset with a subsample designed to match transaction and feature counts in prior research. Source: Neupane et al. 2024a}

\tabularnewline

            \toprule
            \multicolumn{1}{c}{ } & \multicolumn{2}{c}{\makecell[c]{320 Transactions\tnote{\ddag}}} & \multicolumn{2}{c}{3984 Transactions} \\
            \cmidrule(lr){2-3} \cmidrule(lr){4-5}
            & 25 Features & 110 Features & 25 Features & 110 Features \\
            \midrule
            ACC & 83.34 & 89.24 & 98.12 & 99.02 \\
            TNR & 84.67 & 89.59 & 98.19 & 99.06 \\
            PRE & 81.88 & 89.3 & 98.05 & 98.98 \\
            FPR & 18.12 & 10.7 & 1.95 & 1.02 \\
            FNR & 15.2 & 10.82 & 1.8 & 0.93 \\
            TPR & 84.8 & 89.18 & 98.2 & 99.07 \\
            \bottomrule
            \insertTableNotes

\end{longtable}

    \end{ThreePartTable}

Building upon the benchmark model analysis, Table
\ref{tbl-standaloneRFxGBoost} presents results from a separate study by
Neupane and Griva (2024b), some of whom are co-authors of this current
work. These findings provide an additional perspective on the XGBoost
model's classification performance, particularly its ability to
generalize across varying dataset sizes and feature sets. The consistent
high performance of the XGBoost model, evidenced in both Tables
\ref{tbl-rfComparativeConfusionMatrixBenchMarkMethods} and
\ref{tbl-standaloneRFxGBoost}, reinforces its effectiveness in
distinguishing between lawful and unlawful transactions.

In this subsequent study, Neupane and Griva (2024b) directly compared
the benchmark results using an expanded feature set within the US
market. Notably, each of these models demonstrated enhanced performance
compared to the benchmarks, as detailed in Table
\ref{tbl-standaloneRFxGBoost}. For instance, with 25 features and 320
transactions (columns 1 and 2), the models showed an approximate 0.4
percent improvement over the benchmark results. The best-performing
model, presented in column 4 of Table \ref{tbl-standaloneRFxGBoost},
achieved a remarkable overall accuracy of 99.02 percent, with a TPR of
99.07 percent, a TNR of 99.06 percent, a FNR of 0.93 percent, and a FPR
of 1.02 percent. This improved performance can be attributed to factors
such as the availability of a larger dataset and the inclusion of
randomly selected lawful transactions from a larger pool, which provided
the classifier with more diverse and informative data.

The improvement shown in Table \ref{tbl-standaloneRFxGBoost} underscores
the impact of data richness and feature diversity, suggesting that the
model's reliance on key features like Market Beta and Price Book, as
highlighted by the SHAP plot (see
Section~\ref{sec-analysis-interpretation-and-ranking-covariates} and
Section~\ref{sec-analysis-interpretation-and-ranking-features-after-removal-collinearity-Shapley-based}),
is amplified when provided with a more comprehensive dataset. From a
practical perspective, this comparison demonstrates the potential for
significant gains in classification accuracy when leveraging larger,
more diverse datasets, reinforcing the importance of robust data
collection and feature engineering strategies.

\subsection{Interpretation of the ranking of
features}\label{sec-analysis-interpretation-and-ranking-covariates}

\begin{wrapfigure}{r}{0.52\textwidth} 
    \centering
    \includegraphics[width=0.45\textwidth]{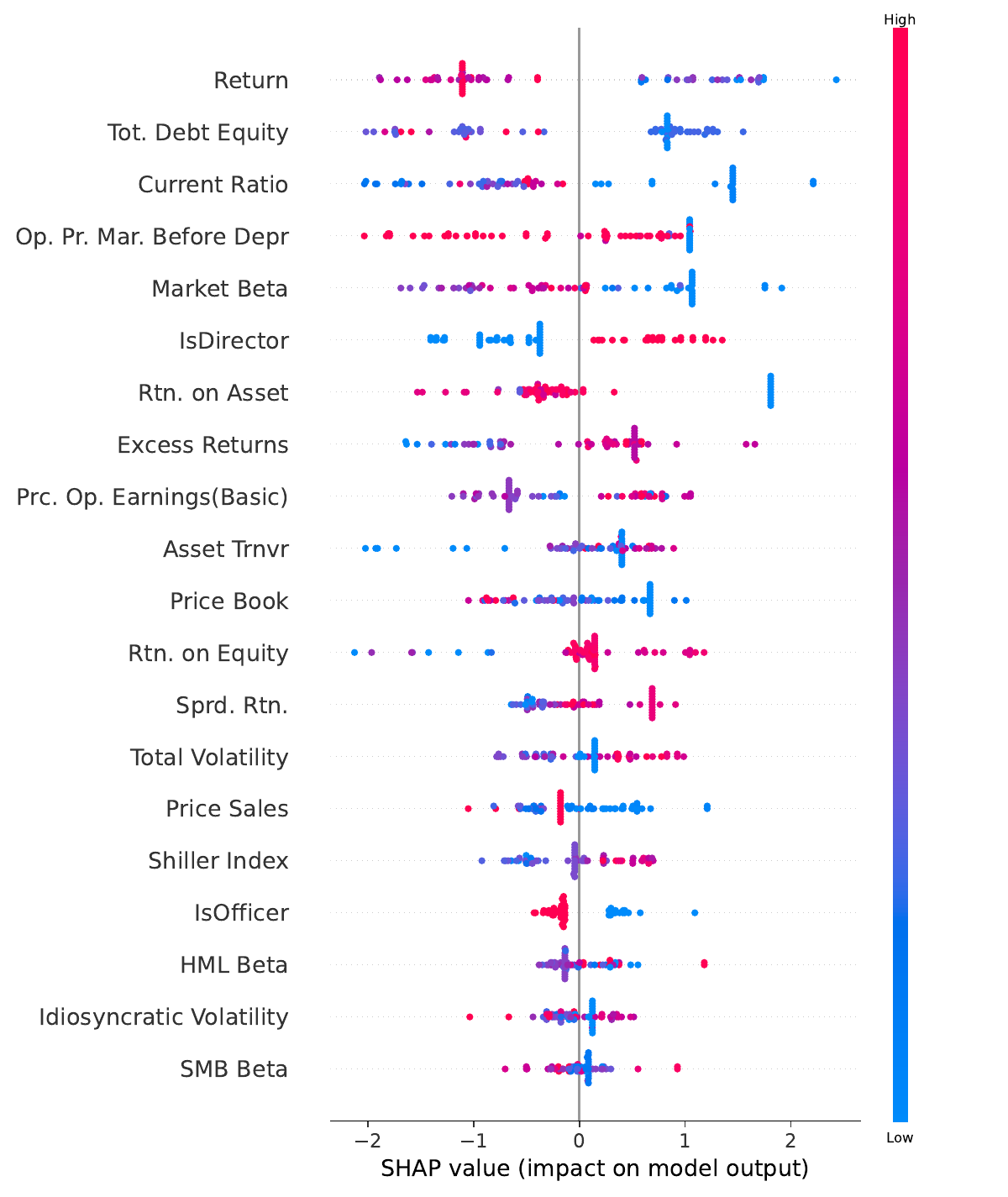}
    \caption{Beeswarm plot illustrating feature importance based on SHAP before feature correlation removal.  Features are ranked in descending order of mean absolute SHAP along the vertical axis. The horizontal axis represents the SHAP, showing the magnitude and direction of each feature's impact on the model's predictions. Dot density along each feature row visualizes the distribution of SHAP. The color bar represents the distribution of raw feature values for each instance (red: high, blue: low), providing insight into the relationship between Shapley and raw values. }  
    \label{fig:varImpMethods xgb_rf_SHAPLEY_before_correl_removal_data}
\end{wrapfigure}

XGBoost's feature importance, derived from split frequency and loss
reduction, offers valuable insights but is susceptible to biases
stemming from correlated features and overfitting, as noted by Strobl et
al. (2008). To mitigate these limitations, this study employs SHAP
values, a game-theoretic approach that provides a nuanced understanding
of feature attribution by quantifying each feature's contribution to
individual predictions (Wang et al. (2024), Ghorbani, Kim, and Zou
(2020)). Higher SHAP values indicate greater influence on predicting
transaction lawfulness, and their distribution reflects both impact and
importance (Lin and Gao (2022), Sun et al. (2020)). This methodological
shift allows for more reliable interpretation, particularly in datasets
with complex feature relationships. As visualized in Figure
\ref{fig:varImpMethods xgb_rf_SHAPLEY_before_correl_removal_data}, the
SHAP beeswarm plot reveals the influence of various features on the
predictive model's output, showcasing their relative importance and
impact direction.

``Return'' as a top ranked feature demonstrates the strongest predictive
power, suggesting companies with specific return profiles are more
likely to be associated with unlawful activities. However, its mixed
impacts could signify complex, non-linear relationships or interaction
effects. ``Total Debt to Equity,'' ``Current Ratio,'' and ``Operating
Profit Margin Before Depreciation'' emerge as highly influential,
indicating that financial health and operational efficiency are key
determinants in classifying transactions. ``Market Beta'' and ``Price
Book'' are influential but ranked lower, with their mixed impacts
potentially reflecting sophisticated insider strategies or, more likely,
confounding effects from correlated features like Total Volatility, SMB
Beta, and Return. ``Price Sales'' exhibits a predominantly negative
impact, suggesting the model's sensitivity to overvaluation. Its
correlation with other valuation metrics (Return on Equity, Operating
Profit Margin) complicates isolating its true effect. ``IsDirector'' and
``IsOfficer'' highlight the role of corporate governance in predicting
unlawful behavior. ``Excess Returns,'' ``Price Operating Earnings,''
``HML Beta,'' ``Idiosyncratic Volatility,'' and ``SMB Beta'' show weak
or inconsistent impacts, suggesting they are less reliable predictors.

A crucial consideration is the potential influence of feature
correlation. As Figure
\ref{fig:varImpMethods xgb_rf_SHAPLEY_before_correl_removal_data} is
generated before correlation removal, SHAP values may not accurately
reflect isolated feature impacts. Correlated features can `share'
importance, overestimating some influences while masking others. For
instance, `Return's' dominance may be partially due to correlated
performance metrics. Therefore, this plot offers a valuable overview but
requires cautious interpretation due to multicollinearity. Analyzing
feature importance post-correlation removal (Figure
\ref{fig:permImp aftermulRemovalPermImp_xgb_shapley_BAR}) provides a
more accurate view of individual contributions. The observed feature
interplay provides context for performance differences between Tables
\ref{tbl-rfComparativeConfusionMatrixBenchMarkMethods} and
\ref{tbl-standaloneRFxGBoost}. The model's high accuracy (84.99 percent
in Table \ref{tbl-rfComparativeConfusionMatrixBenchMarkMethods} and
99.02 percent in Table \ref{tbl-standaloneRFxGBoost}) may be attributed
to its ability to capture relationships between key financial indicators
and unlawful transactions. The substantial improvement with increased
data and features underscores the importance of top-ranked features and
the model's capacity to leverage richer information. However, the
ambiguous impacts of `Price Operating Earnings' and `Return,' along with
the low importance of `HML Beta' and `Acq. Disp.,' indicate potential
areas for refinement. The mixed `Return' and `Price Operating Earnings'
impacts could signify complex relationships or interaction effects. The
low `HML Beta' and `Acq. Disp.' importance may suggest redundancy or
minimal unique contribution. The mixed `Market Beta' impact, and the
negative `Price Book' and `Price Sales' impacts are complicated by
feature correlation. While these patterns could align with certain
unlawful trading scenarios, multicollinearity renders interpretations
unreliable. These SHAP values may reflect the influence of correlated
variables, not isolated insider actions. To address this, iterative
Variance Inflation Factor (VIF) analysis (detailed in
Section~\ref{sec-method-causal-forests}) was applied, confirming that
for the final feature set, all VIF scores were below the conservative
threshold of 10.0. Specifically, Price-to-Book and Market Beta exhibited
low VIF scores of 1.58 and 1.82, respectively. In essence, analyzing
feature importance after correlation removal is crucial for accurate
understanding, though correlation does not equal causation, and further
investigation is always needed.

\subsection{Removal of Correlation between
features}\label{sec-analysis-interpretation-and-ranking-features-before-remove-collinearity}

\begin{figure}
     \centering
     \begin{subfigure}[b]{0.49\textwidth}
         \includegraphics[width=\textwidth]{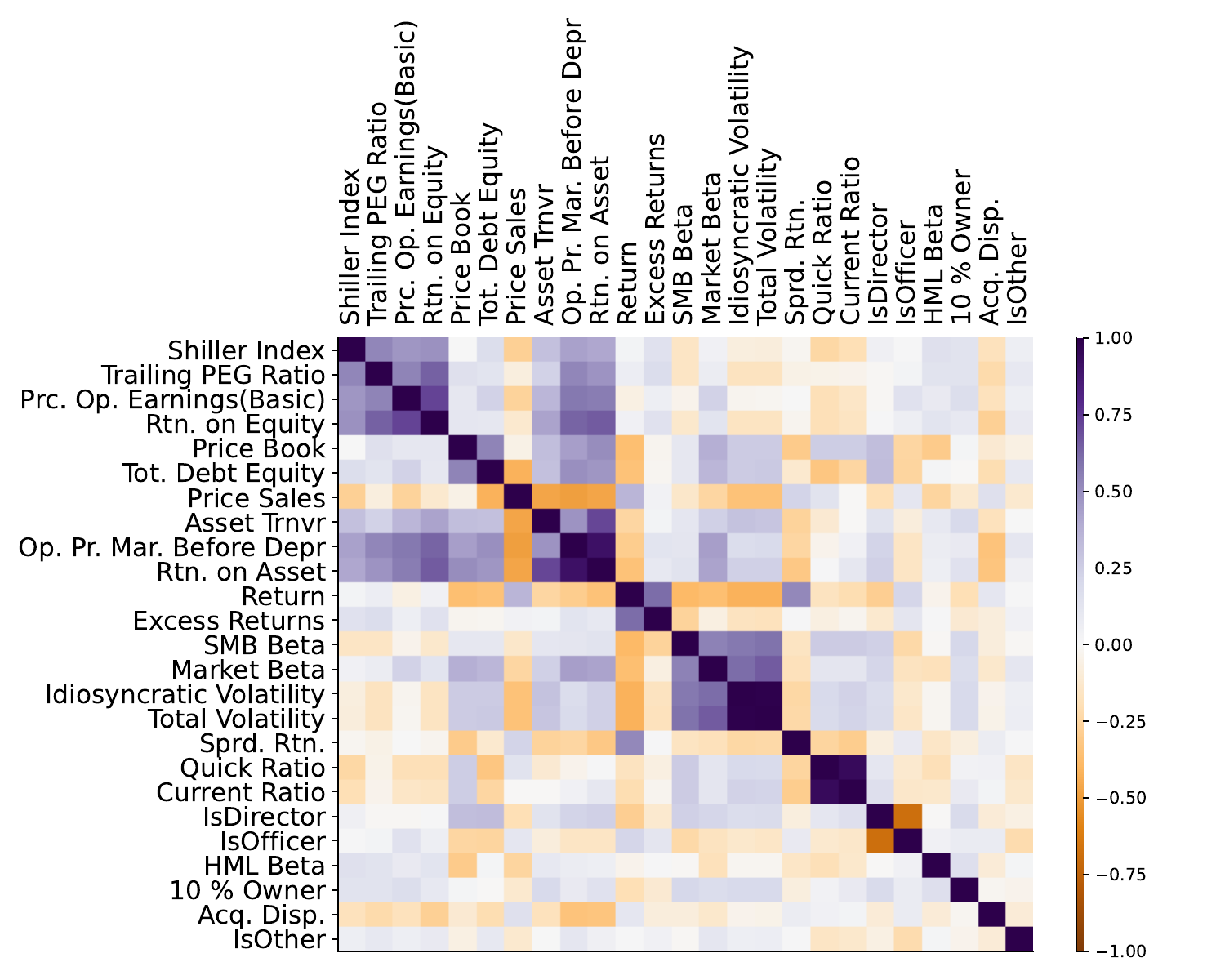}
         \caption{The Heatmap}
         \label{fig:removalCluster_rf_xgb_denogram_heatmap}

     \end{subfigure}
      \begin{subfigure}[b]{0.49\textwidth}
         
         \includegraphics[width=\textwidth]{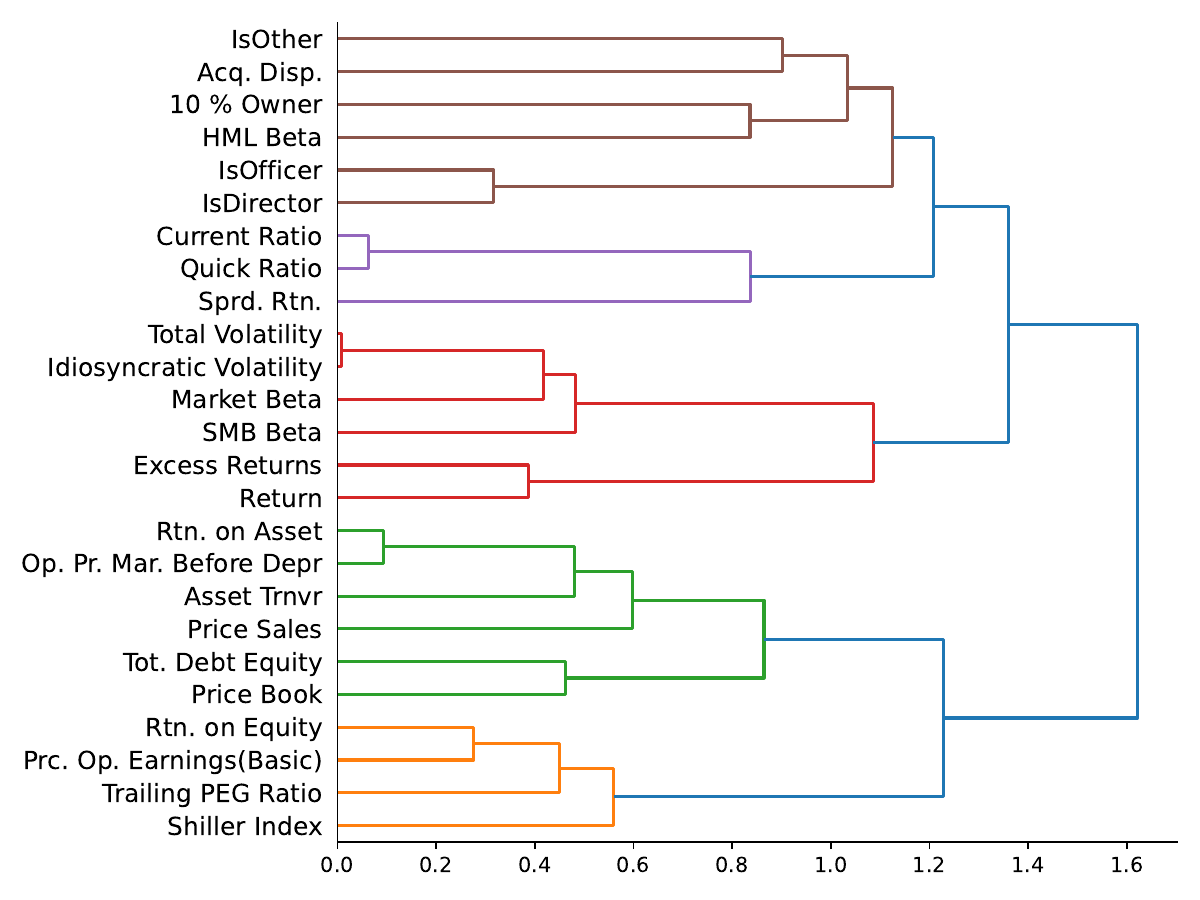}
         \caption{Dendogram representing hierarchial clusters}
         \label{fig:removalCluster_rf_xgb_denogram_correlation_plot}
     \end{subfigure}
     \caption{Hierarchical clustering of the features based on the Spearman rank-order correlations displayed as heatmap in Figure \ref{fig:removalCluster_rf_xgb_denogram_heatmap} and dendogram in Figure \ref{fig:removalCluster_rf_xgb_denogram_correlation_plot}. The Figure \ref{fig:removalCluster_rf_xgb_denogram_heatmap} depicts correlation and clusters based on few selected features shown for illustrative purposes only. Correlation plot shows the strength of association between features which is represented by proportion of the gradient of color ranging between $-1$ (dark purple) and $+1$ (dark orange). The color gradient in each square represent the corresponding coefficient of correlation (\(\rho\)). The figure shows the formation of clusters among the ``similar'' variables. Figure \ref{fig:removalCluster_rf_xgb_denogram_correlation_plot} represents formation of hierarchial clusters between features organized as clades. The sequentially merged pairs of similar features form neighbors to each other. Source: \cite{neupane2024xgboost} }  
      \label{fig:featureImp rf_xgb_dendo__training_data}
\end{figure}

This section discusses the influence of feature correlation and the
method used to address it: hierarchical clustering. A heatmap
\ref{fig:removalCluster_rf_xgb_denogram_heatmap} and a dendrogram
\ref{fig:removalCluster_rf_xgb_denogram_correlation_plot} visually
represent the correlation matrix and hierarchical clustering of
features, respectively. The heatmap
\ref{fig:removalCluster_rf_xgb_denogram_heatmap} displays the Spearman
rank correlation matrix. Darker colors indicate stronger correlations,
with purple/blue for positive and orange/brown for negative
correlations. Distinct blocks of strong correlations are evident, such
as the cluster of ``Price Book,'' ``Price Sales,'' ``Total Debt
Equity,'' ``Return on Equity,'' and ``Asset Turnover'' suggesting shared
underlying financial characteristics. Another cluster includes
``Return,'' ``Excess Returns,'' ``Operating Profit Margin Before
Depreciation.'' and ``Return on Asset'' likely reflecting different
aspects of profitability and performance. ``Market Beta'' and
``Idiosyncratic Volatility'' show negative correlations with some
features in the profitability cluster. The dendrogram
\ref{fig:removalCluster_rf_xgb_denogram_correlation_plot} illustrates
the hierarchical clustering based on correlation distances. Closely
related (highly correlated) features are grouped together on branches
closer to the root. Branch height reflects dissimilarity (greater
distance means less similarity). The dendrogram visually confirms the
clusters observed in the heatmap. For example, ``Price Book,'' ``Price
Sales,'' ``Total Debt to Equity,'' ``Return on Equity,'' and ``Asset
Turnover'' form a tight cluster, as do ``Return,'' ``Excess Returns,''
``Operating profit margin before depreciation'' and ``Rtn. on Asset.''
``IsDirector'' and ``acquisition-disposition'' are closely related to
``HML beta.'' ``Shiller Index,'' ``Trailing PEG Ratio,'' ``Prices
Operating Profit (basic),'' and ``Return on Equity'' form another
cluster.

Strong feature correlations pose a significant challenge for
interpreting feature importance. Correlated features share information,
and if one feature from a correlated group is selected for splitting in
a tree-based model, it can mask the importance of other related
features. The model might arbitrarily choose one representative feature,
overestimating its importance and underestimating the others. This is
particularly problematic for UIT. For example, if ``Return'' and
``Excess Returns'' are highly correlated, and ``Return'' is identified
as a strong predictor, it might not be ``Return'' itself that's
indicative of insider trading, but the shared underlying information.
Using importance derived from correlated data can lead to
misinterpretations. Therefore, accounting for correlation is crucial for
accurate feature importance analysis. By disentangling shared
information, one can assess each feature's unique contribution. After
correlation removal, the previously tight clusters in the dendrogram
would likely spread out, reflecting distinct roles.The hierarchical
clustering in the dendrogram underscores the need for addressing
multicollinearity for reliable feature importance analysis. This allows
for a more accurate assessment of individual importance. For instance,
``Excess Returns'' might emerge as a more significant predictor than
``Return'' after decorrelation. In summary, while the heatmap and
dendrogram visualize feature relationships, strong correlations require
caution in interpreting feature importance.

\subsection{SHAP based ranking of features after removal of
correlation}\label{sec-analysis-interpretation-and-ranking-features-after-removal-collinearity-Shapley-based}

Recognizing the high correlation among financial and trade-related
features, this section uses SHAP to quantify the marginal contribution
of each feature and provides evidence that removal of correlation
significantly enhanced the interpretation. It supports the notion as
suggested by Avanzi et al. (2023), Sigrist (2023) and Meinshausen (2008)
that ranking correlated features can dilute the unique contributions of
individual feature, potentially leading to an underestimation of their
true importance.

\begin{wrapfigure}{r}{0.5\textwidth} 
    \centering
    \includegraphics[width=0.48\textwidth]{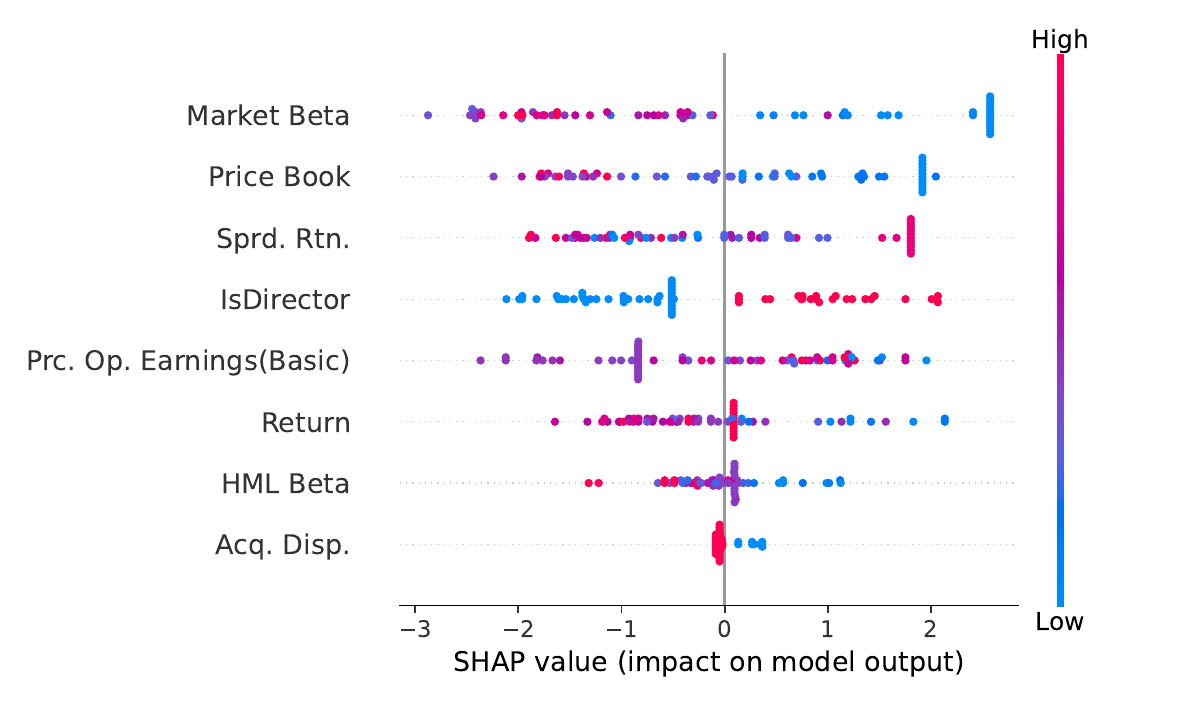}
    \caption{SHAP values for UIT, calculated after removing multicollinearity. This plot shows the independent impact of each feature on UIT prediction, ranked by mean absolute SHAP values. The x-axis represents the range of SHAP values, and color indicates the raw feature value.}  
    \label{fig:permImp aftermulRemovalPermImp_xgb_shapley_BAR}
\end{wrapfigure}

Figure \ref{fig:permImp aftermulRemovalPermImp_xgb_shapley_BAR}
illustrates the SHAP-based feature rankings, showcasing a combination of
trading, governance, and financial metrics that contribute to the
prediction of UIT. To isolate each feature's distinct impact and
mitigate the distorting effects of multicollinearity, hierarchical
clustering was employed to eliminate correlated variables. Features were
retained for further analysis only if their mean absolute SHAP values
exceeded a threshold of 0.022, thereby focusing on those with a
demonstrably substantial influence on the model's predictive outcomes.
Beyond these the SHAP values drops signficantly. This rigorous selection
process yielded eight key features: Market Beta, Price Book Value,
Spread of Return, IsDirector, Price Operating Earnings, Market Return,
High Minus Low Beta, and Acquisition and Dispositions. These refined
features, now free from the confounding influences of multicollinearity,
serve as the basis for subsequent causality testing, aimed at
determining the direction and strength of relationships between these
key indicators and UIT likelihood. The resulting feature set, as
depicted in Figure
\ref{fig:permImp aftermulRemovalPermImp_xgb_shapley_BAR}, offers a more
precise depiction of each feature's unique contribution to the model's
predictive accuracy.

Low Market Beta values (blue dots) exhibit a strong positive association
with transactions classified as unlawful, suggesting that periods of
reduced market volatility are a significant indicator of potential UIT.
This aligns with research on information asymmetry (Akerlof (1970)),
where Akerlof highlighted how information gaps lead to market
inefficiencies that insiders exploit. Models like Kyle (1985)
demonstrate how informed traders can discreetly leverage private
information in less volatile markets. Kyle's model shows how informed
traders strategically trade to maximize profits while minimizing price
impact, a strategy more easily employed in low-volatility environments.
Insiders might adapt their strategies to exploit information advantages
in low-beta stocks (Cohen, Frazzini, and Malloy (2010)), which often
have lower liquidity and less analyst coverage. Cohen's research details
that insiders can leverage their informational advantages particularly
well in less closely monitored stocks. Behavioral finance (Shefrin
(2002)) and agency theory (Jensen and Meckling (2019)) also support
this, with Shefrin noting that overconfidence can drive excessive
trading, and Jensen highlighting how agent self-interest can lead to
information exploitation.

Low Price Book ratios (blue dots) are strongly linked to a higher
likelihood of unlawful transactions, indicating that insiders may target
undervalued companies. This corresponds with the theory of information
asymmetry (Akerlof (1970)), where insiders exploit their knowledge to
capitalize on mispriced assets. This is consistent with empirical
findings that value stocks tend to outperform growth stocks (Fama and
French (1993)), as Fama and French's model explains how value stocks,
often with low PB ratios, offer higher returns due to higher risk. And
that insider trading profitability is often higher in companies with
lower PB ratios (Piotroski (2000)), Piotroski's work shows that value
firms with strong financial signals yield higher returns, which insiders
might exploit.

Spread of Return displays a multifaceted impact on UIT prediction,
showing a mix of red, blue, and purple dots. Higher values (red dots)
generally push towards ``lawful'' classification, but can sometimes
increase the likelihood of ``unlawful'' classification (blue dots),
suggesting a complex, non-linear relationship. Purple dots indicate a
mixed influence. This aligns with behavioral finance and market timing
research (Barber and Odean (2012)), where Barber and Odean discuss how
overconfidence can lead to excessive trading, and ( Lakonishok,
Shleifer, and Vishny (2005), Jeng, Metrick, and Zeckhauser (2003)),
which present mixed findings on insiders' ability to time markets. And
with information asymmetry theories (Easley et al. (1996), Glosten and
Milgrom (1985)), Glosten and Milgrom's work explains how bid-ask spreads
widen with informed trading, while Easley and O'Hara describe how
information-based trading affects liquidity.

IsDirector strongly pushes transactions involving directors (red dots)
towards an ``unlawful'' classification, consistent with agency theory
and information asymmetry. Agency theory (Jensen and Meckling (2019))
posits that directors might act in their self-interest, and information
asymmetry (Akerlof (1970)) suggests they have privileged access to
information.

Price Operating Earnings (Basic) exhibits a multidimensional and dynamic
relationship with UIT, showing a mix of red, blue, and purple dots,
defying simple interpretation. Both high (red dots) and low (blue dots)
values can be associated with unlawful trading, potentially due to
earnings management and information asymmetry (Beneish (1999)),
Beneish's model detects earnings manipulation, and (Shiller (2003)),
Shiller's work challenges efficient market theory, suggesting mispricing
can be exploited. Purple dots indicate a mixed influence. Its weak
influence after correlation removal suggests complex interplay between
earnings information, insider behavior, and market dynamics.

The `Return' demonstrates a complex interplay with the likelihood of UIT
after accounting for multicollinearity. The post-correlation removal
(Figure \ref{fig:permImp aftermulRemovalPermImp_xgb_shapley_BAR}), the
influence of `Return' becomes ambiguous, exhibiting a dispersed pattern
with a mix of red, blue, and purple dots. Specifically, low returns
(blue dots) tend to correlate with an increased probability of UIT,
aligning with theoretical and empirical justifications: theoretically,
insiders possessing non-public negative information may trade prior to
public disclosure, resulting in low pre-transaction returns, or face
financial distress prompting sales. Empirically, this pattern may
reflect cases where insiders sell to mitigate losses based on
foreknowledge of negative news. Conversely, moderate to high returns
(purple/red dots) generally associate with a decreased probability of
UIT, suggesting routine sales or portfolio diversification following
positive stock performance. Empirically, this could capture situations
where insiders sell for legitimate reasons after a stock's appreciation.
However, the non-linear distribution highlights that high returns can
also be linked to UIT, possibly due to insiders capitalizing on
short-term market fluctuations with non-public information. This
suggests that low returns are a stronger indicator of potential UIT than
high returns are of lawful trading, though the overall impact of
`Return' is moderate compared to other features. This variability can be
attributed to factors highlighted in behavioral finance and market
inefficiency literature, such as investor overconfidence (Daniel,
Hirshleifer, and Subrahmanyam (1998)) and market bubbles/crashes
(Shiller (2003)), which can influence the relationship between returns
and insider trading.

HML Beta, like Return, exhibits a dispersed pattern, though with a
narrower distribution of SHAP values. This dispersion, characterized by
a mix of red, blue, and purple dots, indicates that both positive and
negative values of HML Beta influence the prediction of UIT. While the
range of HML Beta's impact is less pronounced than that of Return, its
distribution suggests it captures a specific dimension of trading
behavior. Notably, the presence of blue dots correlating with an
increased probability of UIT implies that low HML Beta values can be
indicative of potential unlawful activity. This finding necessitates a
nuanced understanding of the Fama-French Three-Factor Model (Fama and
French (1993)). Although the model primarily attributes HML's
explanatory power to broad market returns associated with value versus
growth stocks, the SHAP plot demonstrates that HML Beta also captures
aspects of individual stock behavior relevant to UIT predictions.
Specifically, it might be capturing information related to how insiders
trade based on their perception of a stock's value or growth potential,
which could be informed by non-public information. Therefore, HML Beta's
influence, while moderate, is not insignificant. It likely captures a
specific dimension of trading behavior related to value/growth factors
that are relevant to the prediction of UIT, warranting further
investigation.

Acquisition and Disposition exhibits minimal influence, showing a tight
cluster of dots around zero, indicating that individual purchase and
sale activities are not strong independent predictors of UIT. This could
be due to the diversity of individual motivations and the subtlety of
insider trading patterns (Cohen, Malloy, and Pomorski (2012), Cohen,
Frazzini, and Malloy (2008)), Cohen's research highlights the difficulty
of detecting insider trading patterns. Behavioral theories (Kahneman and
Tversky (1979)) further highlight the complexity of individual
decision-making, as Kahneman and Tversky's work shows how individuals
deviate from rational economic behavior.

The methodology effectively isolated the independent contributions of
financial and governance features in predicting UIT by employing SHAP
values and eliminating multicollinearity. This post-correlation
analysis, detailed in Figure
\ref{fig:permImp aftermulRemovalPermImp_xgb_shapley_BAR}, provides a
more precise understanding of each feature's impact and substantiates
the high classification accuracy achieved in our models (Table
\ref{tbl-standaloneRFxGBoost}). Market Beta emerged as the dominant
predictor, underscoring the crucial role of market volatility and
information asymmetry. Price Book, Spread Return, and IsDirector also
proved significant, emphasizing the importance of valuation metrics,
market microstructure, and corporate governance. Return, Price Operating
Earnings, HML Beta, and Acquisition and Disposition exhibited weaker or
minimal influence, suggesting their limited independent predictive
power. The consistent patterns observed in the SHAP plots align with
established financial and behavioral theories, validating the model's
ability to capture meaningful relationships within the data. This
multi-faceted approach, combining high classification accuracy with
detailed feature importance analyses, provides robust evidence for the
model's validity and its effectiveness in identifying and predicting
UIT. The removal of multicollinearity, coupled with SHAP values, yielded
a more accurate and interpretable model, enhancing our understanding of
the factors contributing to UIT.

\subsection{Do certain proxies have a direct effect on
UIT?}\label{sec-analysis-causal-analysis-ranking-features-causal-forest}

This section explores the complex problem of the UIT detection by
integrating predictive modeling with causal inference. While XGBoost
achieves high predictive accuracy, relying solely on its predictive
power has limitations, particularly concerning interpretability and
addressing uncertainty. SHAP values, while insightful, also have
limitations in fully capturing the complexities of causal relationships.
To overcome these limitations, this study employs a CF model, building
upon the predictive capabilities of XGBoost. The CF model estimates
CATE, quantifying the average difference in potential outcomes between
treated (top eight features) and untreated features, providing a deeper
understanding of the causal impact of these ``treatments'' on UIT
classification.

\begin{wrapfigure}{r}{0.5\textwidth} 
    \centering
    \includegraphics[width=0.48\textwidth]{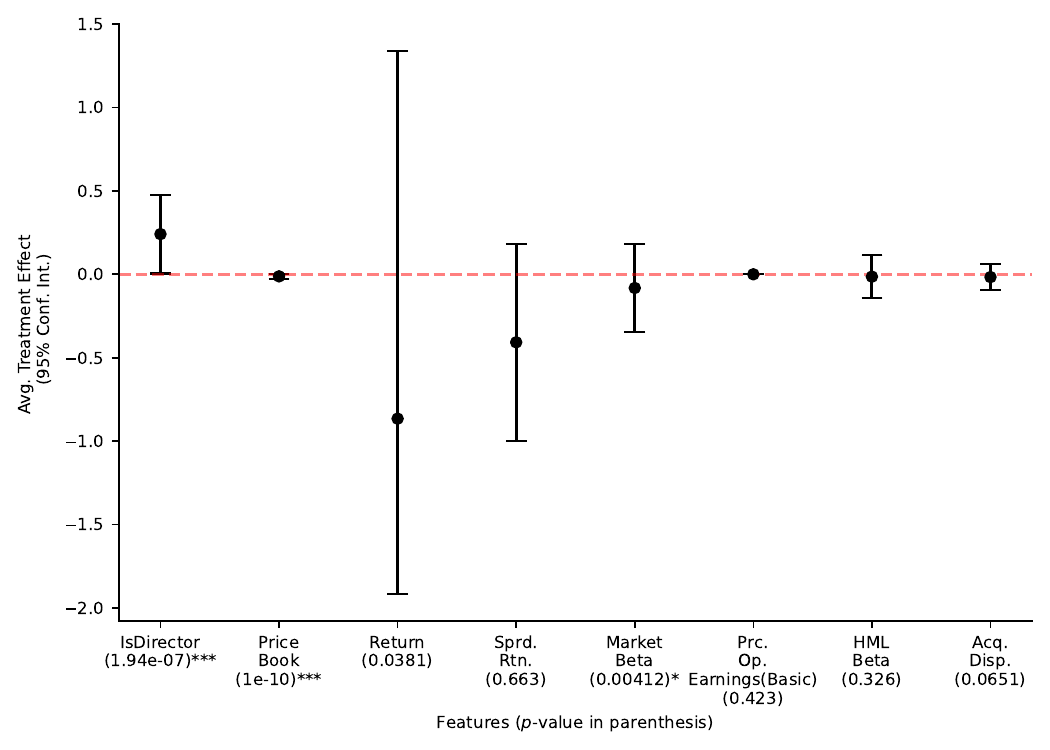}
    \caption{Confidence Interval ($95$ precent) of the selected predictive features (treatments) ordered by their SHAP value based relative importance (see Figure \ref{fig:permImp aftermulRemovalPermImp_xgb_shapley_BAR}). The error bar represent if samples are drawn randomly but repeatedy $95$ precent of the CI would contain the unknown population mean within upper and lower limits. The causal significance (***) is measured by $p$-value $\leq 0.05$ at the $95$ percent.}  
      \label{fig:pValRanking afterRemovalofCorrelation}
\end{wrapfigure}

The causal analysis, stemming from the analysis of feature importance
for potential UIT, reveals several key insights when comparing SHAP
values before and after correlation removal with the ATE. Generally, a
strong correlation is observed between the ranking of features based on
their SHAP values and their causal significance, as detailed in Figure
\ref{fig:pValRanking afterRemovalofCorrelation}. Following SHAP-based
feature selection and correlation removal, the top eight features were
utilized as treatment variables in a CF model, which estimated their
causal impact on UIT. Figure
\ref{fig:pValRanking afterRemovalofCorrelation} ranks these features by
p-values, derived from an XGBoost model, illustrating their causal
impact on UIT.

Figure \ref{fig:pValRanking afterRemovalofCorrelation} provides
compelling causal evidence for the heightened risk of insider trading
associated with directorships. The ATE plot reveals a statistically
significant positive ATE for IsDirector (p \textless{} 0.001),
indicating that holding a director position directly increases the
likelihood of engaging in UIT. This causal link reinforces the patterns
observed in the SHAP plots (Figures
\ref{fig:varImpMethods xgb_rf_SHAPLEY_before_correl_removal_data} and
\ref{fig:permImp aftermulRemovalPermImp_xgb_shapley_BAR}), where being a
director consistently shows a positive association with transactions
classified as unlawful. This convergence of evidence highlights the
importance of considering both correlational and causal analyses when
investigating UIT.

The analysis of Price-to-Book (PB) ratio across the three figures (SHAP
plots: Figures
\ref{fig:varImpMethods xgb_rf_SHAPLEY_before_correl_removal_data} and
\ref{fig:permImp aftermulRemovalPermImp_xgb_shapley_BAR}; ATE plot:
Figure \ref{fig:pValRanking afterRemovalofCorrelation}) reveals a
consistent and compelling narrative about its role in UIT, particularly
highlighting the importance of causal inference alongside predictive
modeling. Both SHAP plots demonstrate a clear pattern: lower PB ratios
are associated with a higher likelihood of transactions being classified
as unlawful, suggesting that insiders may be drawn to undervalued
companies. This observation is further strengthened by the causal
analysis in Figure \ref{fig:pValRanking afterRemovalofCorrelation},
which shows a statistically significant negative ATE for PB, indicating
that higher PB ratios actually decrease the likelihood of insider
trading.

Figure \ref{fig:pValRanking afterRemovalofCorrelation} highlights a
statistically significant causal effect of `Return' on UIT, with a
p-value of 0.0381, below the standard 0.05 significance level. The ATE
for `Return' is -8.64e-01, indicating that higher past returns are
linked to a lower probability of UIT. This negative ATE suggests that
higher past returns would decrease UIT likelihood, counter to initial
expectations. The discrepancy between this negative ATE and the mixed or
weakly positive influence observed in the SHAP plots underscores the
importance of employing causal inference methods to disentangle the true
impact of variables on UIT. This intricate relationship may be tied to
risk-taking behavior, as insiders might become more cautious and less
likely to engage in UIT during periods of low returns.

The ATE for Spread Return is close to zero, but the narrow confidence
interval indicates a statistically significant causal effect.
Specifically, the negative ATE suggests that higher Spread Return leads
to a decrease in the likelihood of UIT. This implies that wider spreads
might deter insider activity, potentially due to increased scrutiny or
higher transaction costs associated with wider spreads. The causal
evidence from Figure \ref{fig:pValRanking afterRemovalofCorrelation}
enriches the intricate interplay between Spread Return and UIT revealed
in the SHAP plots. In both SHAP plots, Spread Return exhibits a mixed
influence, where higher values generally push towards classifying
transactions as ``lawful,'' but can sometimes increase the likelihood of
``unlawful'' classification. This mixed influence could be due to
behavioral factors, such as overconfidence and market timing attempts by
insiders, as discussed earlier.

While the SHAP plots initially suggest a positive association between
low beta and UIT, the ATE plot presents a contrasting perspective. The
ATE plot provides causal evidence that increasing Market Beta has a
statistically significant negative effect on the likelihood of UIT. This
counterintuitive finding implies that higher market volatility may
actually discourage UIT, potentially due to heightened market scrutiny
or increased liquidity. This observation aligns with the theory that
insiders may prefer less volatile environments for their unlawful
activities.

Figure \ref{fig:pValRanking afterRemovalofCorrelation} reveals that Prc.
Op. Earnings (Basic) does not exhibit a statistically significant causal
effect on IT. The ATE is near zero, and the wide confidence interval,
along with a p-value of 0.423, confirms the absence of a reliable causal
link. This contrasts with the mixed distribution observed in the SHAP
plots, where both high and low values of Prc. Op. Earnings (Basic)
appear to influence the model's predictions, suggesting a complex,
potentially non-linear relationship. This discrepancy underscores the
difference between correlational and causal analyses.

Features such as `HML Beta,' and `Acquisition and Disposition' exhibited
statistically insignificant ATE at the conventional 0.05 level. This
aligns with their limited influence observed in the SHAP plots,
suggesting a weaker direct causal relationship with UIT after accounting
for feature correlation. After mitigating the influence of
multicollinearity, the analysis demonstrates a strong alignment between
feature importance as determined by SHAP values and causal significance
as assessed by p-values from the ATE analysis. This alignment reinforces
the robustness of the identified relationships, indicating that features
deemed highly predictive by SHAP often correspond to statistically
significant causal effects (low p-values). Conversely, features with
higher p-values suggest that their observed effects may be attributable
to random variation.

The analysis consistently demonstrates the value of integrating causal
inference with predictive modeling in the study of UIT. The XGBoost
model's selection of `Return,' `Market Beta,' `IsDirector,' `Spread of
Return,' `Price to Operating Earnings,' `Price Book/Sales,' and
`Acquisition and Disposition' as key predictive features is supported by
both Shapley rankings and the Average Treatment Effect (ATE)
significance. While discrepancies arise, notably with `Market Beta,'
`Price Book,' and `Return' exhibiting counterintuitive negative ATEs,
these differences highlight the impact of model-specific functional
forms and the presence of confounding variables. The CF analysis, by
mitigating these confounding effects, provides more reliable insights
into the direct causal impacts of these features, particularly
`IsDirector,' `Price Book,' and `Market Beta,' which emerge as key areas
of interest. This confirms that the removal of correlation and
subsequent ATE analysis offer a more robust understanding of UIT
compared to Shapley plots based on correlated data. The consistent
identification of these features reinforces their strong association
with UIT and validates the effectiveness of combining ensemble methods
with causal inference techniques, even as further validation remains
essential to navigate the inherent challenges of causal analysis.

\section{Discussions}\label{sec-analysis-discussions}

The research found that a limited set of influential features
significantly explain UIT. To validate this, transactions were
classified, features were ranked by their SHAP, and conducted a causal
analysis by treating the top eight features as ``treatments'' and the
remaining as controls. While SHAP identify predictive features, causal
analysis determines their true impact, differentiating correlation from
causation, especially in scenarios with correlated features. The top
eight features, aligned with the financial anomalies literature listed
by (\textbf{hou2017replicating?}), are crucial for UIT classification
and suggest that insiders exploit information for financial gain. The
results confirm that institutional, financial, and trading features
drive UIT.

Connecting these ATE findings with the classifier performance data from
Table~\ref{tbl-standaloneRFxGBoost} provides further context. The high
accuracy and TNR achieved model in detecting UIT highlight the
effectiveness of machine learning in this domain. A high TNR is
particularly crucial, as it minimizes false accusations. XGBoost's
performance stems from its ability to capture complex, non-linear
relationships within the data, which is essential when dealing with
intricate financial market dynamics. The high accuracy of XGBoost
classification indirectly contributes to the credibility of both SHAP
and CF analysis. Another indirect proof of the credibility is
consistency of results of both SHAP and CF analysis. Shifting most
important feature values up or down moves both SHAP and CF model
predictions towards lawful or unlawful conclusions in a consistent way.
The significant performance gains observed in
Table~\ref{tbl-standaloneRFxGBoost} with larger datasets and more
features underscore the importance of data availability and feature
engineering. The fact that the ATE plot identifies ``IsDirector,''
``Price Book,'' and ``Market Beta'' as influential features suggests
that the classifiers are likely leveraging the information contained
within these variables to distinguish between lawful and unlawful
transactions. The strong positive ATE of ``IsDirector,'' coupled with
the high accuracy and TNR of the models, raises concerns about potential
information asymmetry and warrants further investigation. The
counterintuitive negative ATE of ``Market Beta,'' especially after
correlation removal, is a particularly intriguing finding. It suggests
that the model is picking up on a complex relationship between market
risk and insider trading, potentially revealing strategies that exploit
market stability. This nuanced understanding of market dynamics,
combined with the ability to leverage key features like ``IsDirector''
and ``Price Book,'' likely contributes to the high performance observed
in the classification tasks. For an instance, as being director explains
the UIT which is consistent to Goergen, Renneboog, and Zhao (2019). The
privileged positions allow directors to become ``opportunistic traders''
and ``have the most predictability for future firm events'' and
therefore be able to generate profit from the ``returns'' as an
opportunistic trader that lasts at least for ``six months'' (Cohen,
Malloy, and Pomorski (2012)). Proxies that contain price information
such as returns, market beta, book/price are used by insiders to exploit
asset mis-pricing and are good predictors to detect the UIT (Ahern
(2018), Hirshleifer (2001)). Similarly, insiders ramped up their
acquisition and disposition when price-to-book ratio is less than
\(1.0\) for substantial gains. Such results point to the fact that
``positive abnormal returns following insider purchases and negative
abnormal returns following insider sales'' (Reinganum (1988)). The
quarterly rise in transactions the price increases experienced a
substantial price rises (Banz (1981), Seyhun (1986)). This phenomenon is
associated with the ``noise trading'' when other ordinary investors
follow the insiders in purchase and sales (Fishman and Hagerty (1995)).
Consistent with literature, the observations suggest that causal
discovery aligns to human reaction that they evaluate events based on
causal chains, assign significance based on outcomes (Sloman (2009)). In
fact, by onboarding CF the results are able to close the gap produced by
uncertainty due to only SHAP because the results of SHAP and causally
significant features are almost similar. Hence, the method exploited
causal knowledge to extract contribution and assign importance to
features. The mechanism yields intuitive explanations relatable to the
precise discovery of structures of causal attributes in data to predict
the target variable (Breuer et al. (2024)).

\section{Conclusions and Future Work}\label{sec-conclusions-future}

This research demonstrates that effectively identifying, interpreting,
and explaining the UIT, a complex real-world problem, necessitates a
deep understanding of causality. The resulting model exhibits strong
classification accuracy, along with robust feature ranking (using
Shapley) and causal significance (using CF), facilitating the discovery
of unique causal relationships. This involves considering alternative
scenarios and evaluating their potential outcomes. Within a
high-dimensional feature space, the proposed architecture integrated
state-of-the-art techniques. By integrating advanced machine learning
techniques, including XGBoost, SHAP, and CFs, this study achieves higher
classification accuracy and provides valuable insights into both feature
importance and causal relationships. The analysis reveals several key
findings: ``IsDirector'' emerges as a strong positive indicator of
potential UIT, suggesting a link between directorship and advantageous
trading outcomes. ``Price Book'' exhibits a significant negative impact,
potentially reflecting insiders acting on knowledge of overvaluation.
Most notably, ``Market Beta'' demonstrates a counterintuitive negative
relationship with the outcome variable, even after accounting for
feature correlation. This suggests a more nuanced understanding of
market risk by insiders, possibly related to profiting from market
stability. While these findings are statistically significant, the wide
confidence intervals observed for some features underscore the need for
cautious interpretation. Decisions based on these features, especially
those with confidence intervals spanning zero, should be made with
careful consideration of the inherent uncertainty.

Moving forward, several avenues of research can enhance the robustness
and applicability of this framework. First, expanding the dataset size,
exploring alternative model specifications, and incorporating additional
control variables could help reduce uncertainty and refine the estimates
of causal effects. Validating the findings with alternative
methodologies or additional data sources is also crucial for confirming
the observed relationships. Second, reframing the UIT problem as a
multi-class classification task, encompassing a broader range of illicit
trading activities, could provide a more comprehensive and realistic
representation of the issue, leading to more effective regulatory
policies. This multi-class perspective aligns with human cognition and
provides crucial insights for more effective regulatory policy
formulation (Athey (2019)). Third, significantly expanding the feature
space to include a wider array of potential predictors, such as the
numerous asset pricing anomalies identified in prior research (for
example, (\textbf{hou2017replicating?})), could further improve the
model's power and reliability. Finally, investigating the influence of
partially confounded settings on feature contributions to causal
structures and their impact on Shapley value rankings is a promising
direction for future research. This could shed light on how unobserved
confounding factors might affect the interpretation of feature
importance and causal effects in the context of IT detection. By
pursuing these directions, future research can further refine our
understanding of UIT and develop more powerful tools for its detection
and prevention.

\section*{Appendix}\label{sec-appendix}
\addcontentsline{toc}{section}{Appendix}

\newpage
\section*{Appendix}\label{sec-appendix}
\addcontentsline{toc}{section}{Appendix}

\% Set the caption before the longtable {[}cite: 2025-11-20{]}
\begingroup \small 

\begin{longtable}{p{3.5cm} p{11.5cm}}

\caption{\label{tbl-Variables}List of features used in the study (* represent matching to the previous study)}

\tabularnewline

  \\
\toprule
\textbf{Group} & \textbf{Variables} \\
\midrule
\endfirsthead

\multicolumn{2}{c}%
{{\bfseries \tablename\ \thetable{} -- continued from previous page}} \\
\toprule
\textbf{Group} & \textbf{Variables} \\
\midrule
\endhead

\midrule
\multicolumn{2}{r}{{Continued on next page...}} \\
\endfoot

\bottomrule
\endlastfoot

Activity/Efficiency Ratios & Asset Turnover*, Inventory Turnover, Payables Turnover, Receivables/Current Assets \\ \addlinespace

Annual Valuation Ratios & Shiller's P/E, Dividend Yield, Dividend Payout Ratio, Enterprise Value Multiple, Price-to-Cash Flow, Price-to-Earnings, excl. EI (diluted)*, Price-to-Earnings, incl. EI (diluted)*, Forward P/E to 1-year Growth (PEG) ratio*, Forward P/E to Long-term Growth (PEG) ratio*, Trailing PEG Ratio, Price-to-Sales Ratio* \\ \addlinespace

Capitalization Ratios & Capitalization Ratio, Long-term Debt/Invested Capital, Common Equity/Invested Capital, Total Debt/Invested Capital \\ \addlinespace

Financial Soundness Ratios & Cash Flow to Debt, Cash balance to Total Liabilities, Current Liabilities as Percentage of Total Liabilities, Total Debt as percentage of Total Assets, Gross debt to EBITDA, Long-term Debt/Book Equity, Free Cash Flow/Operating Cash Flow, Interest as Percentage of Average Long-term Debt, Interest as Percentage of Average Total Debt, Inventory/Current Assets, Long-term Debt/Total Liabilities, Total Liabilities/Total Tangible Assets, Operating Cash Flow to Current Liabilities, Profit before D\&A to current liabilities, Receivables Turnover, Short-Term Debt/Total Debt \\ \addlinespace

Liquidity Ratios & Cash Conversion Cycle, Cash Ratio, Current Ratio*, Quick Ratio (Acid Test)*, Quoted Spread \\ \addlinespace

Miscellaneous Ratios & Accruals/Average Assets, Advertising as Percent of Sales, Market Capitalization, Price-to-Book Ratio*, Research and Development as percent of Sales, Sales per Dollar Total Stockholders’ equity, Sales per Dollar Invested Capital, Sales per Dollar Working Capital, Labor Expenses/Sales \\ \addlinespace

Ownership/Governance & Acquisition Disposition, Derivatives Held, Adjusted Derivatives Held, IsDirector, IsOfficer, IsOther, Ten Percent Ownership \\ \addlinespace

Profitability Ratios & After Tax Return on Average Common Equity, After Tax Return on Total Stock Holder's Equity, After Tax Return on Invested Capital, Alpha (Excess Return), Cash Flow Margin, Effective Tax Rate, Trailing PEG Ratio, Gross Profit Margin*, Gross Profit/Total Assets, Net Profit Margin, Operating Profit Margin After Depreciation*, Operating Profit Margin Before Depreciation*, Pre-tax Return on Total Earning Assets, Pre-tax return on Net Operating Assets, Pretax Profit Margin, Return on Assets*, Return on Capital Employed, Return on Equity* \\ \addlinespace

Risk & Ask, Ask or High Price, Beta (High Minus Low)*, Market Beta*, Small-minus-big Size factor*, Bid, Bid Ask Spread, Bid or Low Price, Effective Spread, Excess Return from Risk Model*, Idiosyncratic volatility from the q-factor model, Kyle Lambda, Number of Derivatives, Number of Derivatives after Trade, Number of Trades, Price Impact, Market R-Squared, Realized Spread, Return, Returns without Dividend, Underlying Market Equity Volume, Underlying Shares Adjust, Outstanding Shares, Underlying Market Price, Underlying Market Price Adjust, Spread of Return, Total Volatility*, Volume, Exercise Price, Exercise Price Adjust \\ \addlinespace

Solvency Ratios & Debt-to-equity Ratio, Debt-to-assets, Debt-to-Capital, After Tax Interest Coverage, Interest Coverage Ratio \\ \addlinespace

Valuation Ratios & Price-to-Operating EPS, excl. EI (basic), Price-to-Operating EPS, excl. EI (diluted) \\ 

\end{longtable}

\endgroup

\section{Compliance with Ethical
Standards}\label{compliance-with-ethical-standards}

Funding This research received no external funding or financial
assistance during its preparation.

Competing Interests The author certify that they have no conflicts of
interest, financial or otherwise, to disclose.

Author's Declaration on AI Assistance

The author bear sole responsibility for all substantive ideas and
analyses within this manuscript. Portions of the text were reviewed for
language, style, and clarity through AI-assisted copy editing,
specifically using a large language model (LLM). No autonomous content
creation was performed by the LLM

\section*{References}\label{references}
\addcontentsline{toc}{section}{References}

\phantomsection\label{refs}

\end{document}